\documentclass[aps,pra,twocolumn,nofootinbib,groupedaddress,floatfix]{revtex4-2}

\usepackage{graphicx}
\usepackage{amssymb}
\usepackage{amsmath}
\usepackage{amsfonts}
\usepackage{multirow}
\usepackage{longtable}

\begin{document}

\title{
Enhancement of parity-violating energy difference 
of CHFClBr, CHFClI, and CHFBrI
by breaking the cancellation among valence orbital contributions 
}

\author{Naoya Kuroda}
\affiliation{Department of Micro Engineering, Kyoto University, Kyoto 615-8540, Japan}
\author{Ayaki Sunaga}
\altaffiliation{Present address : Department of Physics, Graduate School of Science, Kyoto
University, Kyoto 606-8502, Japan}
\affiliation{Institute for Integrated Radiation and Nuclear Science, Kyoto University, Osaka 590-0494, Japan}
\author{Masato Senami}
\email{senami@me.kyoto-u.ac.jp}
\affiliation{Department of Micro Engineering, Kyoto University, Kyoto 615-8540, Japan}

\date{\today}

\begin{abstract}
The enhancement of the parity-violating energy difference (PVED) by electronic excitation
is studied for H$_2X_2$ ($X =$ O, S, Se, Te), CHFClBr, CHFClI, and CHFBrI.
To clarify the enhancement mechanism,
the dihedral angle dependence of the PVED of 
H$_2X_2$ in excited states is studied.
If the contribution from the highest occupied molecular orbital (HOMO) to the PVED in the ground state
is larger than the sum of those from all occupied orbitals,
the PVED in the first excited state has a much larger value compared to the ground state
due to cancellation breaking among valence orbital contributions.
This enhancement is named cancellation breaking enhancement.
The PVED enhancement is also studied for CHFClBr, CHFClI, and CHFBrI in excited states,
and 
the cancellation breaking enhancement is confirmed.
When the PVED contribution from the HOMO
is larger than any other contribution,
the cancellation breaking enhancement hypothesis
provides the estimate of PVED in the first excited state
from the HOMO contribution.

\end{abstract}

\maketitle
\section{Introduction}

Parity-violating energy difference (PVED) 
is the energy difference between an enantiomeric pair,
which is induced by the weak interaction between nuclei and electrons in a molecule.
PVED is studied by many groups in experimental and theoretical fields
and is one subject related to
the origin of homochirality in nature \cite{text:homochirality}.
In addition,
if PVED is experimentally observed,
it is the first confirmation of the imprint of the weak interaction on molecules.
A subsequent comparison of the measured PVED value with theoretical prediction 
may constrain new physics beyond the standard model in particle physics,
since many extended models have new particles contributing to the PVED.
Hence, PVED may provide a new tool 
wherein molecules are used to investigate new physics beyond the standard model.

Thus far, 
PVED has not been experimentally observed,
because of its small size.
The tightest upper bound comes from the experiment using the CHFClBr molecule
\cite{CHFClBr}.
In this experiment, 
the vibrational frequency difference between the enantiomeric pair of the molecule
was measured and $|\Delta \nu / \nu| < 2.5 \times 10^{-13}$ was reported.
This bound constrains the PVED with the relation, 
$E_{\rm el}^{\rm PV} / E_{\rm el} 
\sim
E_{\rm vib}^{\rm PV} / E_{\rm vib} $~\cite{Letokhov}.

The usage of molecules with large PVED,
in addition to the improvement of experimental technology,
is important for the observation of nonzero PVED in future experiments.
The PVED of chiral molecules is predicted 
by quantum chemistry computations with relativistic effects.
Many works\cite{Wormit:2014, CHFBrI, Rhenium} have reported larger PVED values than CHFClBr,
and we consider that the search for molecules with larger PVED is still important.
It is known that molecules with heavy elements have large PVED values 
due to their large spin-orbit coupling,
but computations of the PVED of these molecules require huge costs.
It is impossible to exhaustively compute the PVED of molecules with heavy elements.
Hence, we take another route to hunt for molecules with large PVED:
usage of the electronic excitation of chiral molecules.

In Ref.~\cite{Senami:2019},
it was proposed that 
the PVED of electronic excited or ionic states of a chiral molecule
is significantly enhanced compared to the neutral ground state.
The contributions to the PVED from valence orbitals
are canceled out 
in the ground states of many chiral molecules,
and 
the PVED is expected to be enhanced significantly
if this cancellation is broken by excitation or ionization.
In our previous paper \cite{Kuroda:2022},
this speculation was confirmed 
for H$_2X_2$ ($X=$O, S, Se, Te) 
by Coupled Cluster Singles and Doubles (CCSD) and 
Equation-of-Motion CCSD computations.
In the work, 
it was reported that 
the PVED enhancement in the optimized H$_2X_2$ structure
is much larger than the structure with $\phi = 45^\circ$
where $\phi$ is the dihedral angle of H$_2X_2$.
In the present paper,
the PVED enhancement mechanism is studied in detail
through the dependence on the dihedral angle.
In addition, 
we confirm that this enhancement occurs in other molecules:
CHFClBr, CHFClI, and CHFBrI.
CHFClBr was used in the observational experiment \cite{CHFClBr}.
CHFClI and CHFBrI are its derivatives
that were studied as candidates for experimental improvement
\cite{CHFBrI,Darquie:2010}.

This paper is organized as follows.
In the next section, 
the definition of the PVED is introduced briefly.
Then,
our computational method and details are described in Sec.~\ref{sec:comp}.
In Sec.~\ref{sec:results}, 
our results are presented.
The dependence of the PVED on the dihedral angle of H$_2 X_2$ molecules is studied
and one mechanism of PVED enhancement is proposed.
Then, PVED enhancement by electronic excitation
is demonstrated for CHFClBr, CHFClI, and CHFBrI,
and our hypothesis of the enhancement mechanism is confirmed for these molecules.
The last section is devoted to our conclusion and discussion.

\section{Theory}

The PVED is defined as twice the parity-violating energy, $ E_{\rm PV} $,
\begin{align}
\Delta E_{\rm PV} = 2 | E_{\rm PV} |.
\end{align}
The dominant part of the PVED is calculated by the following equation,
\begin{align}
E_{\rm PV} = \frac{G_F}{2 \sqrt {2} } \sum_n g_V^n M_{\rm PV}^n ,
\label{Eq:EPV}
\end{align}
where $G_F = 1.663788(6) \times 10^{-5} (\hbar c)^3$ GeV$^{-2}$ is the Fermi coupling constant \cite{PDG}.
The coupling of the nucleus, $n$, is given as $g_V^n = Z^n (1 - 4 \sin^2 \theta_W) - N^n$,
where $Z^n $ and $N^n$ are the number of protons and neutrons in the nucleus $n$, respectively,
and $\theta_W$ is the weak-mixing angle, $\sin^2 \theta_W = 0.23121(4) $ \cite{PDG}.
The definition of $M_{\rm PV}^n $ is the electron chirality at the nucleus $n$,
\begin{align}
M_{\rm PV}^n
=
\int d^3x 
\langle \Psi | \hat \psi_e^\dagger (x) \gamma_5 \hat \psi_e (x) \hat \psi_n^\dagger (x) \hat \psi_n (x) | \Psi \rangle,
\label{eq:MPV}
\end{align}
where $ \hat \psi_e $ and $ \hat \psi_n $ are the field operators of the electron and the nucleus $n$, respectively,
$ | \Psi \rangle $ represents the state vector,
and $\gamma_5$ is defined with gamma matrices, $\gamma_5 = \gamma^5 = i \gamma^0 \gamma^1 \gamma^2 \gamma^3 $.
The ordinary Dirac representation \cite{textbook1} is adopted in this article.
In the derivation of this equation,
protons and neutrons are assumed to have the same distribution in nuclei.
Since the distribution of nuclear density is strongly localized,
$M_{\rm PV}^n$ is almost equal to the electron chirality at the nucleus position,
where the electron chirality density is defined as
$ \langle \Psi | \hat \psi_e^\dagger (x) \gamma_5 \hat \psi_e (x) | \Psi \rangle  $.
The derivation of Eq. (\ref{eq:MPV}) is explained in detail in our previous paper \cite{Kuroda:2022}.

In the current study,
a contribution to $ M_{\rm PV}^n $ from the $i$-th orbital 
is used and is defined as
\begin{align}
M_{{\rm PV},i}^n
=
\int d^3x 
\langle \Psi | \hat \psi_{e,i}^\dagger (x) \gamma_5 \hat \psi_{e,i} (x) 
\hat \psi_n^\dagger (x) \hat \psi_n (x) | \Psi \rangle ,
\label{eq:MPVi}
\end{align}
where $\hat \psi_{e,i}$ corresponds to the electron in the $i$-th orbital.
In this expression, two degrees of the Kramers pair are separately treated,
and $ M_{{\rm PV}}^n = 2 \sum_i M_{{\rm PV},i}^n $ in closed-shell molecules.

\section{Computational Detail}
\label{sec:comp}

The expectation values of $E_{\rm{PV}}$ in the excited states are calculated 
using the Finite-Field Perturbation Theory (FFPT)
\cite{Pople:1968, Pawlowski:2015, Norman:2018} with the EOM-CCSD method \cite{Shee:2018}.
For confirmation of our FFPT computations,
the values of $E_{\rm{PV}}$ in the ground states of our target molecules are 
compared with values based on the Z-vector method in CCSD computations \cite{Shee:2016}.

In our computations,
$E_{\rm PV}$ is calculated as the sum of 
contributions from nuclei, 
$E_{\rm PV} = \sum_n E_{\rm PV}^n $
where $ E_{\rm PV}^n $ is the contribution to $ E_{\rm PV} $ from the nucleus $n$.
In the FFPT method, $E_{\rm PV}^n$ is calculated by the perturbation method
whose perturbation Hamiltonian is chosen to be 
$ H_P^n = \lambda^n 
\int d^3x 
\hat \psi_{e}^\dagger (x) \gamma_5 \hat \psi_{e} (x) \hat \psi_n^\dagger (x) \hat \psi_n (x) $,
where $\lambda^n $ is the perturbation parameter for the nucleus $n$,
and $ \langle \Psi | H_P^n | \Psi \rangle = \lambda^n M^n_{\rm PV}$.
The Hellmann-Feynman theorem tells us the following relation:
\begin{align}
\left. \frac{ \partial E^n \left( \lambda^n \right) }{\partial \lambda^n } \right|_{\lambda^n = 0}
&=
\left\langle {\Psi \left| {\frac{{\partial H}}{{\partial \lambda^n }}} \right|\Psi } \right\rangle 
= M_{PV}^n , 
\label{eq:HFT}
\end{align}
where $H$ is the total Hamiltonian.
The computation of the derivative $\partial E^n / \partial \lambda^n |_{\lambda^n = 0} $ in Eq.~(\ref{eq:HFT})
is performed using the finite difference method
as $\partial E^n / \partial \lambda^n |_{\lambda^n = 0} = (E^n (\lambda^n) - E^n(-\lambda^n) )/ (2\lambda^n) $.
For this computation, the choice of the value of $\lambda^n$ is important. 
For the accurate calculation of the derivative in the finite difference method,
small $\lambda^n$ is suitable for avoiding contamination by the effect of second or higher order derivatives,
while with too small $\lambda^n$, 
the effect of $H_P^n$ is less than computational error.
Hence, an appropriate value of  $\lambda^n$ should be adopted.
The determination of the $\lambda^n$ value is based on a comparison 
between the FFPT and Z-vector results in the ground state.
For excited states, the same values of $\lambda^n $ are used.
Additionally, we have confirmed that 
the values of $E_{\rm PV} $ in excited states
are independent of $\lambda^n $ around the appropriate value of $\lambda^n$.
For the computation of $E_{\rm PV}$,
the contribution from hydrogen atoms is neglected,
since the coupling $g_V^n \simeq 0.0752$ is much smaller than other atoms
and $M_{\rm PV}^H $ is small due to the smallness of the spin-orbit interaction.

Our targets for the $E_{\rm PV} $ computations are
H$_2 X_2$ ($X=$ O, S, Se, and Te), CHFClBr, CHFClI, and CHFBrI molecules.
The structures of H$_2 X_2$ are the same as those in our previous paper \cite{Kuroda:2022}
except for the dihedral angle $\phi$,
which is treated as a free parameter in the present work.
The structures of CHFClBr, CHFClI, and CHFBrI are determined 
by geometrical optimization computations with GAUSSIAN16 \cite{GAUSSIAN16}.
Density functional theory (DFT)
with Becke's three-parameter Lee-Yang-Parr hybrid functional (B3LYP) \cite{Becke, LYP}
is used 
and 
D95V basis sets \cite{D95V} with the Stuttgart-Dresden relativistic effective core potential (Br and I) \cite{SDD}
are adopted.
The structures of excited states 
are chosen to be the same as the ground states for all molecules.

The evaluation of Eq.~(\ref{eq:MPV}) requires relativistic four-component wave functions for electrons, 
and the DIRAC program is used for these electronic structure computations.
For computations of H$_2 X_2$ molecules, the DIRAC19 program \cite{DIRAC19, DIRAC:2020} is used,
while DIRAC21 \cite{DIRAC:2020, DIRAC21} is used for CHFClBr, CHFClI, and CHFBrI.
For the self-consistent field (SCF) method computations,
the Dirac-Coulomb-Gaunt Hamiltonian is adopted,
while molecular mean-field approximations \cite{Sikkema:2009} to the Dirac-Coulomb-Gaunt Hamiltonian 
are adopted for post-SCF computations, i.e., CCSD and EOM-CCSD computations.
For the latter computations,
the RELCCSD module is used \cite{Visscher:1995, Visscher:1996, Shee:2018}. 

In computations of H$_2 X_2$ molecules,
dyall.acv3z \cite{basis_set} is employed
and this choice was confirmed to be sufficiently accurate in our previous paper \cite{Kuroda:2022}.
In CCSD and EOM-CCSD computations of H$_2$O$_2$,
all electrons are treated as correlating orbitals
and all virtual orbitals in active space are taken into account.
In computations of H$_2$S$_2$, H$_2$Se$_2$, and H$_2$Te$_2$,
correlating orbitals are chosen to be 
$2s2p$, $4s4p$, and $5s5p$, respectively,
and 
the cutoff of virtual orbitals in the active space
is 100 Hartree (H$_2$S$_2$ and H$_2$Se$_2$)
and 70 Hartree (H$_2$Te$_2$).
The convergence criterion for EOM-CCSD computations is chosen to be $10^{-10}$
as used in our previous work \cite{Kuroda:2022}.
The distribution of the electron chirality density 
is calculated with the QEDynamics program package \cite{QEDynamics}
for the wave function computed by the Hartree-Fock (HF) method of DIRAC19.

In the computations of CHFClBr, CHFClI, and CHFBrI molecules
the dyall.acv2z basis set \cite{basis_set} is adopted,
and the difference in $E_{\rm PV}$ among basis sets is discussed in a later section.
In the CCSD computations,
the cutoff of virtual orbitals in the active space
is 50 Hartree, 
and 
correlating orbitals are chosen to be 
3d4s4p(Br), 3s3p(Cl), 2s2p(F), 2s2p(C), 1s(H), and 4d5s5p(I) for any molecules.
In EOM-CCSD computations of these molecules,
the convergence criterion is loosened compared to that of H$_2X_2$
due to the limit of our computational resources.
The threshold values of the convergence of CHFClBr, CHFClI, and CHFBrI 
are $10^{-5}$, $10^{-6}$, and $10^{-7}$, or better, respectively.
The validity of our computations with loose criteria
will be discussed later.

For the distribution of nuclei,
the Gaussian distribution functions are assumed \cite{Visscher_Gauss:1997}.
This distribution is used for 
both electronic structure computations
and the evaluation of Eq.~(\ref{eq:MPV}).

\section{Results}
\label{sec:results}

In the following,
the atomic units are employed
and in particular $E_{\rm PV} $ is expressed in Hartree ($E_h$).

\subsection{H$_2 X_2$}
\label{sec:H2X2}

\begin{figure*}[t]
\vspace{-10cm}
  \hspace{-1.5cm}
  \begin{minipage}{.46\linewidth}
	\centering
	\includegraphics[width=12cm, bb = 0 0 612 792]{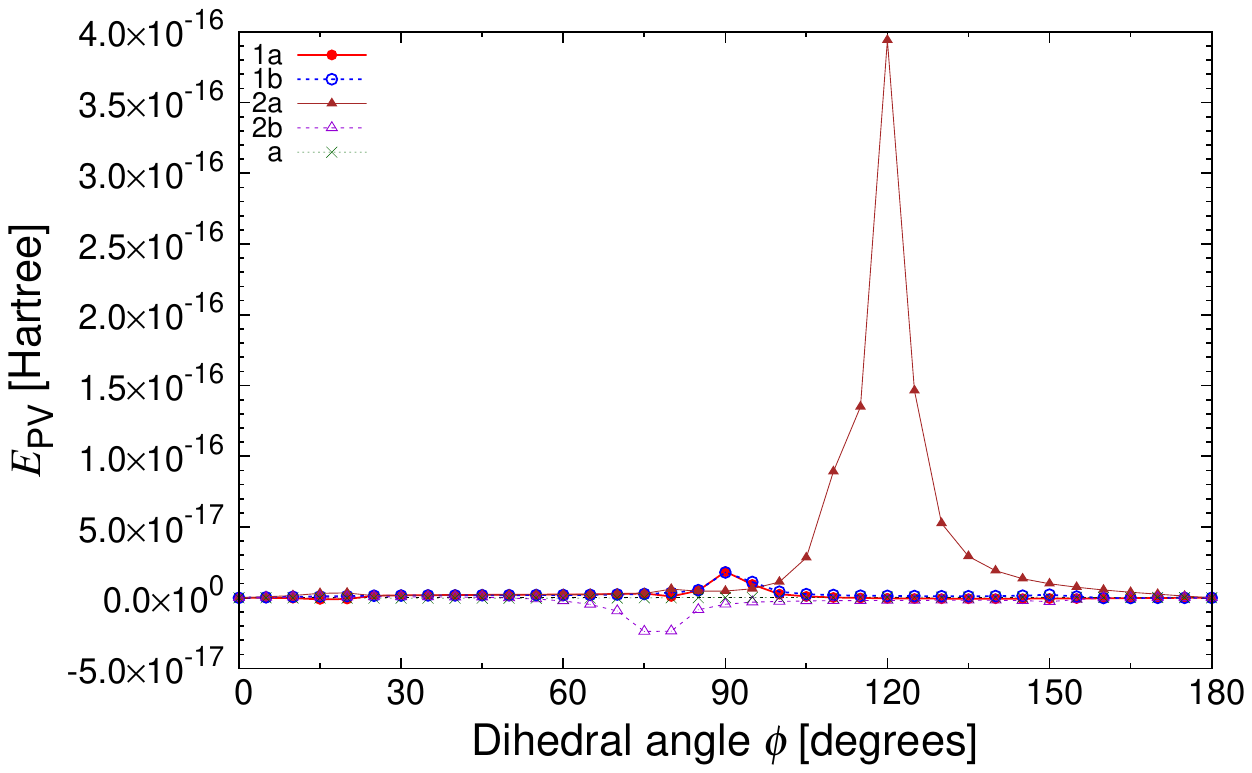} 
 \put(-220,15){ (a) $\rm H_2 O_2$ }
\put(-290,57){ \includegraphics[width=5cm]{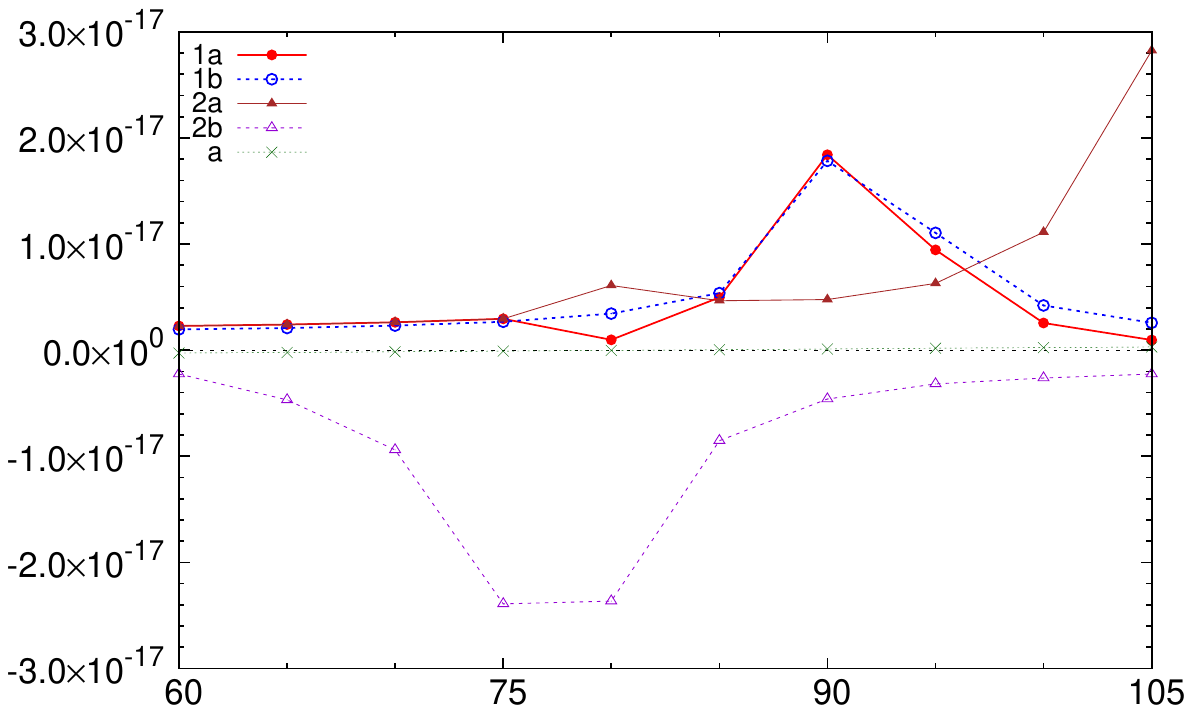} }
 \end{minipage}
	\centering
  \begin{minipage}{.46\linewidth}
\hspace{-.5cm}
	\includegraphics[width=12cm, bb = 0 0 612 792]{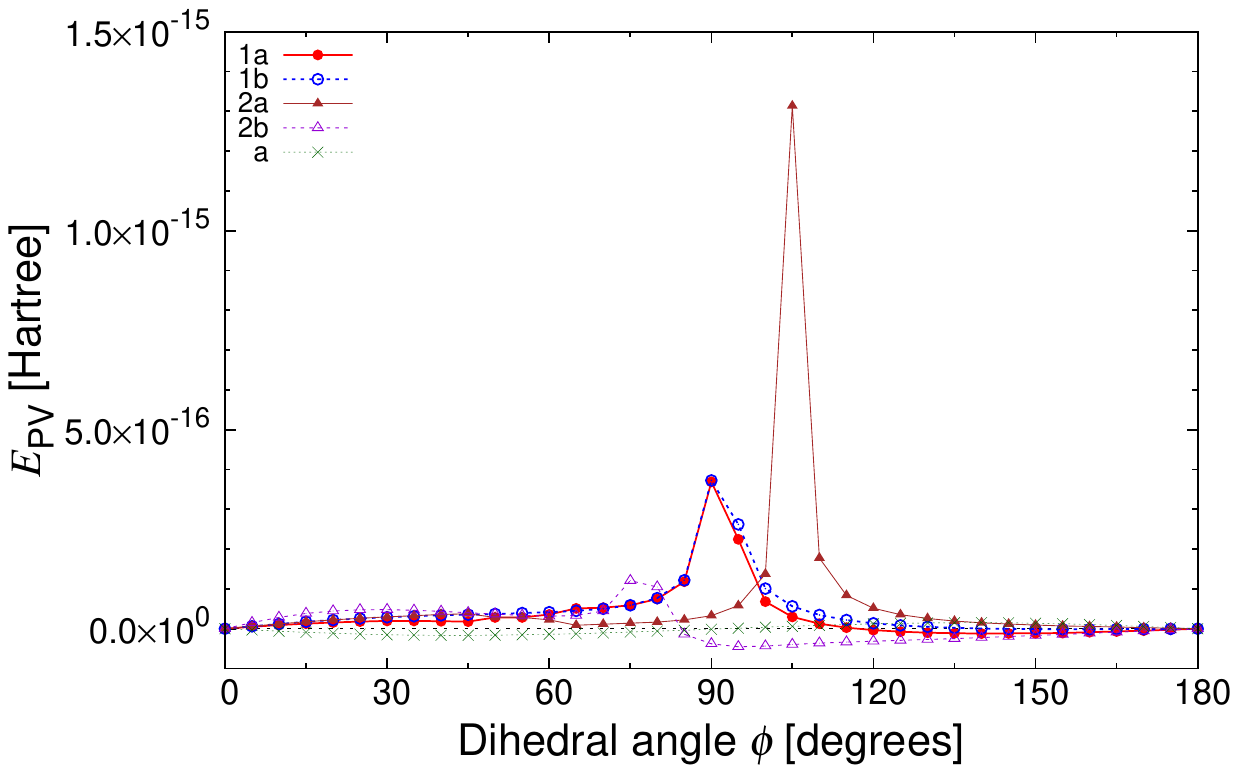} 
 \put(-220,15){ (b) $\rm H_2 S_2$ }
  \end{minipage}
\\
  \hspace{-1.5cm}
  \begin{minipage}{.46\linewidth}
\vspace{-10cm}
	\centering
	\includegraphics[width=12cm, bb = 0 0 612 792]{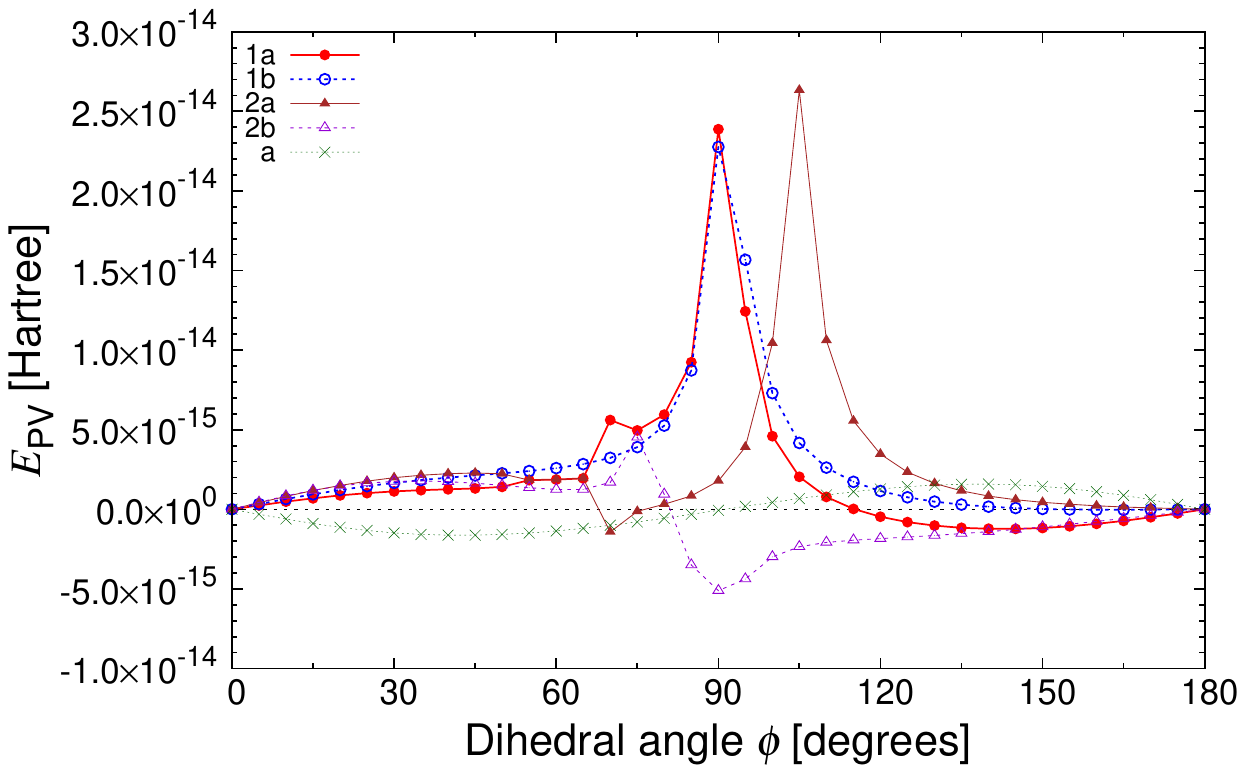} 
 \put(-220,15){ (c) $\rm H_2 Se_2$ }
  \end{minipage}
	\centering
  \begin{minipage}{.46\linewidth}
\vspace{-10cm}
\hspace{-.5cm}
	\includegraphics[width=12cm, bb = 0 0 612 792]{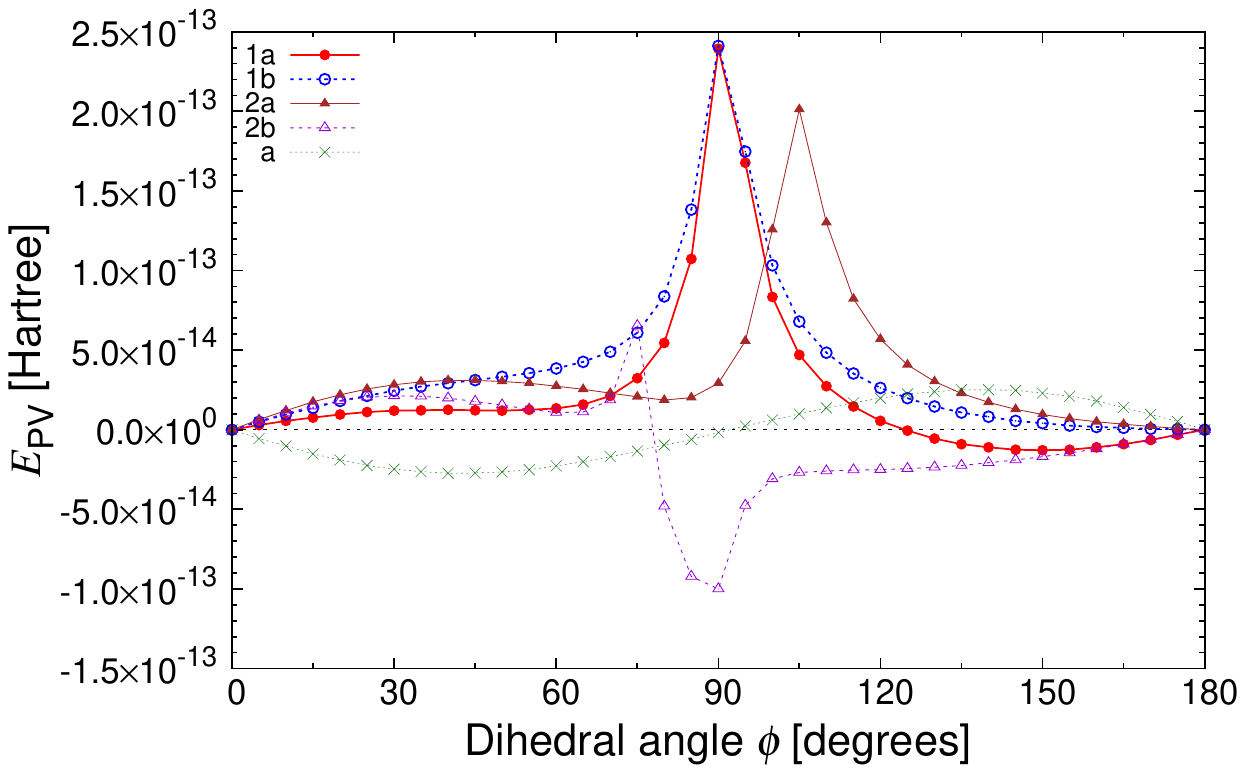} 
 \put(-220,15){ (d) $\rm H_2 Te_2$ }
  \end{minipage}
	\caption{$E_{\rm PV}$ in H$_2X_2$ as a function of dihedral angle $\phi$.
The inset of panel (a) show a magnified view around $\phi = 90^\circ$.
} 
	\label{fig:H2X2_EPV}
\end{figure*}

In our previous paper \cite{Kuroda:2022}, 
the magnitude of the enhancement of $E_{\rm PV}$ in H$_2 X_2$
is different between the optimized and $\phi = 45 ^\circ$ structures.
Hence, 
H$_2 X_2$ is chosen as a model 
and 
the dependence of the enhancement on the dihedral angle is investigated.
In this calculation,
the perturbation parameter $\lambda^n$ in the FFPT method
is set to $10^{-3}$, $10^{-3}$, $10^{-4}$, and $10^{-5}$ (a.u.),
for O, S, Se, and Te, respectively,
which are determined in our previous paper \cite{Kuroda:2022}
by comparing $E_{\rm PV}$ values in the ground state 
with those calculated by the Z-vector method in CCSD computations. 

In Fig.~\ref{fig:H2X2_EPV},
the dependence of $E_{\rm PV}$ on the dihedral angle is shown.
The enhancement is strongly dependent on the dihedral angle
and $E_{\rm PV}$ of H$_2X_2$ exhibits some common peaks.
For the excited states of 1a and 1b,
there is a peak with positive $E_{\rm PV}$ at $\phi = 90^\circ$.
For the 2a state, the positive peak is found at $\phi = 120^\circ $ ($X$=O) and $105^\circ$ ($X$=S, Se, Te).
For the 2b state of H$_2$O$_2$, a negative bump is seen around $\phi = 75^\circ$,
while for H$_2$S$_2$, H$_2$Se$_2$, and H$_2$Te$_2$
the positive peak and negative bump
are observed at $\phi = 75^\circ $ and $\phi = 90^\circ$, respectively.
Since the optimized dihedral angle of H$_2X_2$ 
is $\phi = 115^\circ $ ($X$=O) and $\phi = 90^\circ$ ($X$=S, Se, Te),
large enhancement occurs in the optimized structure.
In this calculation of $E_{\rm PV}$,
the value of $\lambda^n$ is fixed for all values of $\phi$.
This treatment is sufficiently accurate for our study.
The deviation of $E_{\rm PV}$ from the Z-vector result
is within a few percent except for $\phi = 90^\circ$,
where the deviation is 1-7$\%$ due to the smallness of $E_{\rm PV}$ itself.

\begin{figure*}[t]
\vspace{-10cm}
  \hspace{-1.5cm}
  \begin{minipage}{.46\linewidth}
	\centering
	\includegraphics[width=12cm, bb = 0 0 612 792]{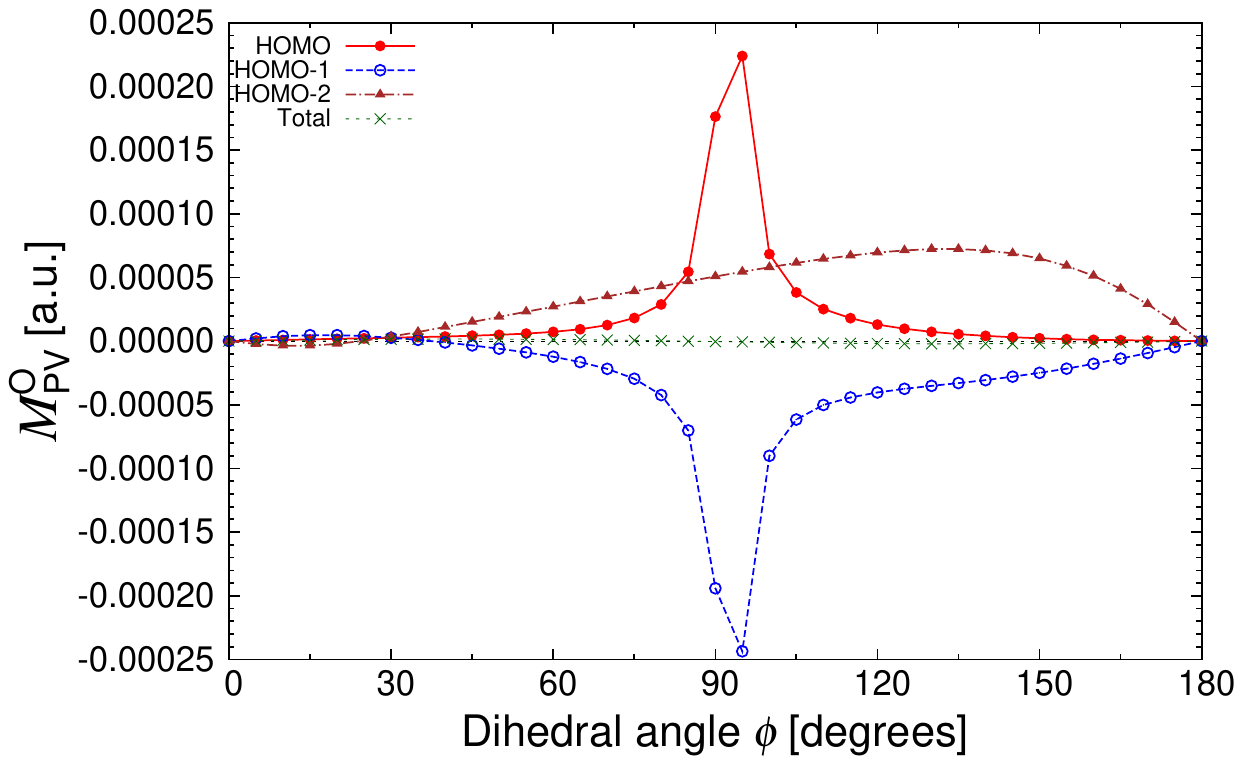} 
 \put(-220,15){ (a) $\rm H_2 O_2$ }
  \end{minipage}
  \hspace{-.5cm}
	\centering
  \begin{minipage}{.46\linewidth}
	\includegraphics[width=12cm, bb = 0 0 612 792]{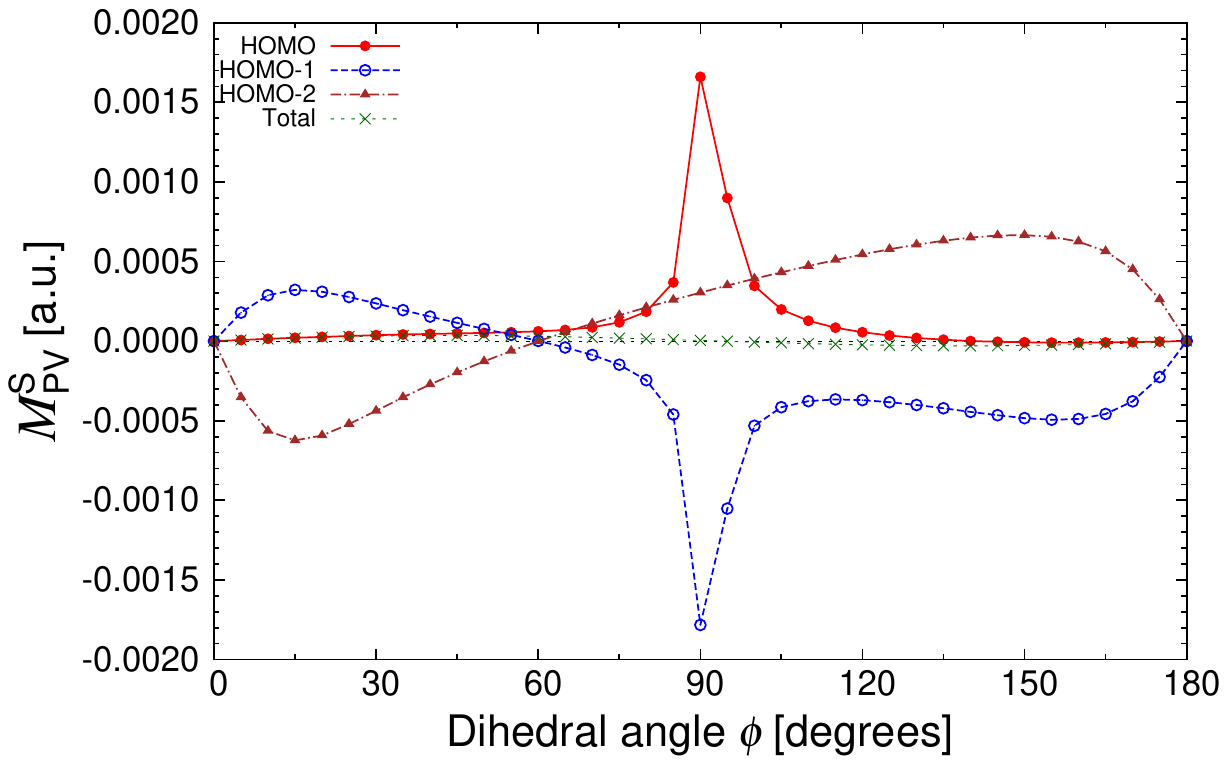} 
 \put(-220,15){ (b) $\rm H_2 S_2$ }
  \end{minipage}
 \\
\vspace{-10cm}
	\centering
  \hspace{-1.5cm}
  \begin{minipage}{.46\linewidth}
	\includegraphics[width=12cm,  bb = 0 0 612 792]{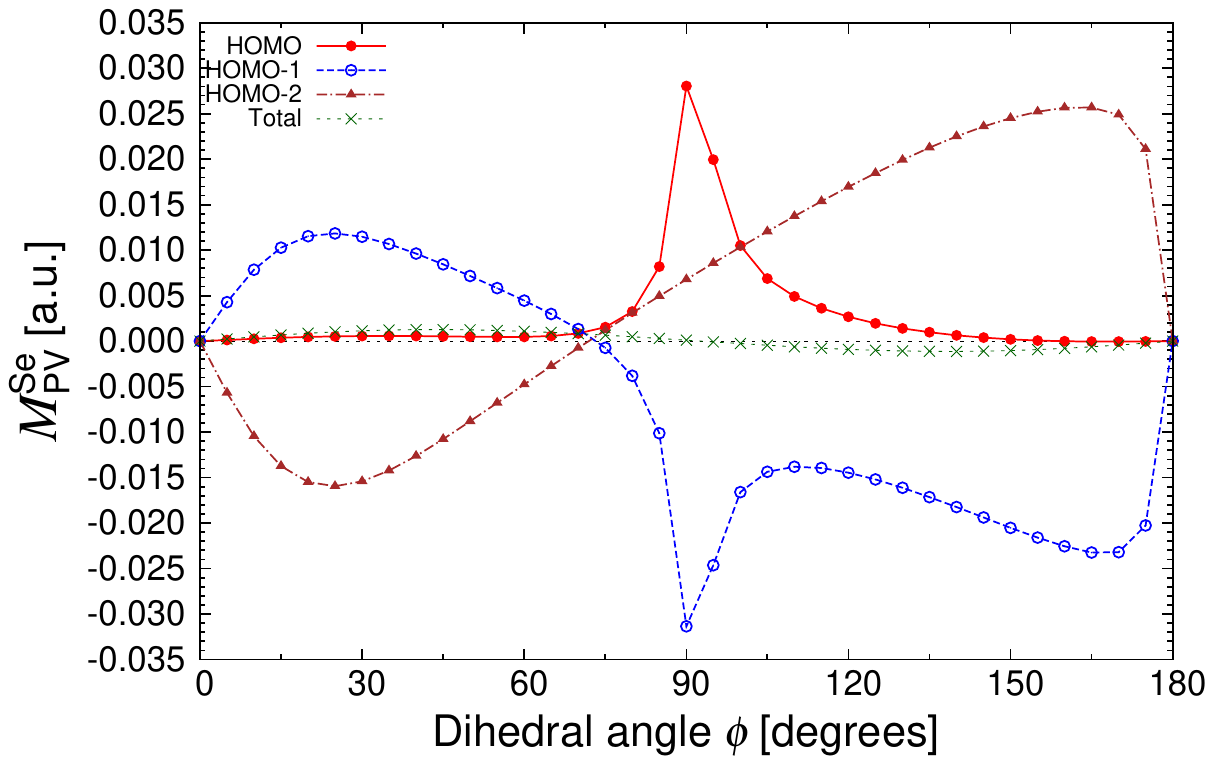} 
 \put(-220,15){ (c) $\rm H_2 Se_2$ }
  \end{minipage}
  \hspace{-.5cm}
	\centering
  \begin{minipage}{.46\linewidth}
	\includegraphics[width=12cm, bb = 0 0 612 792]{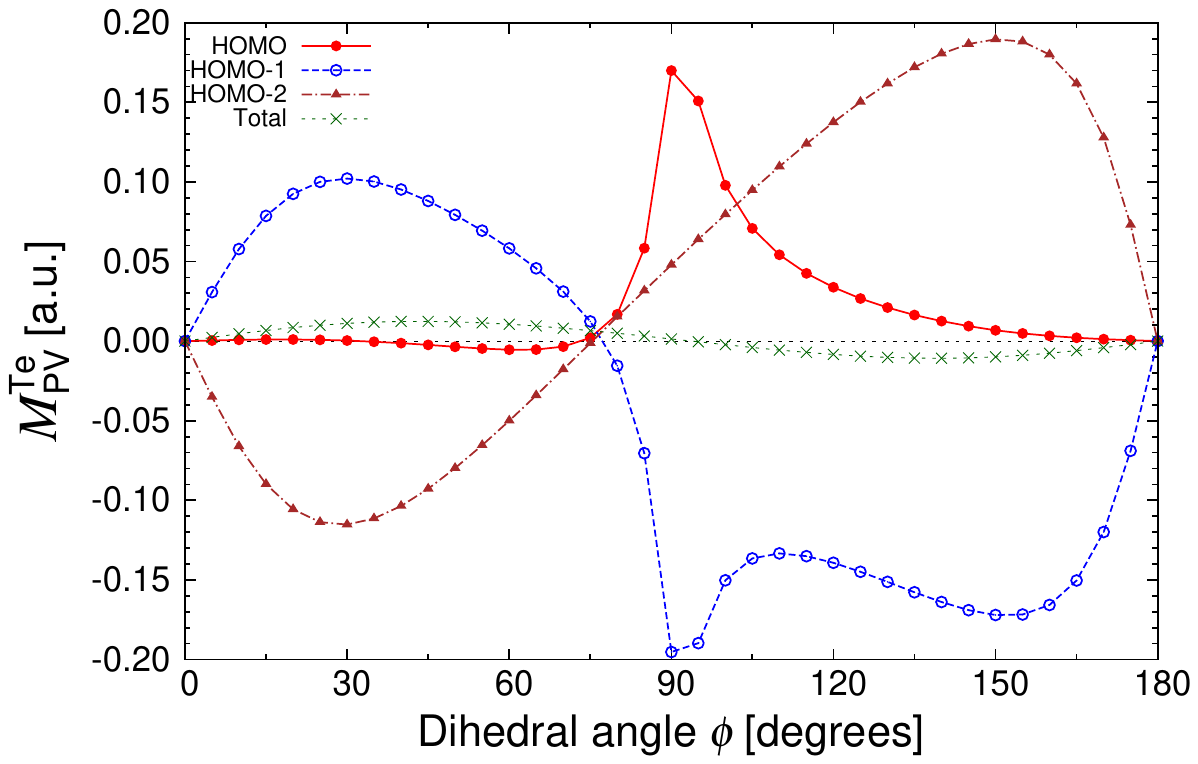} 
 \put(-220,15){ (d) $\rm H_2 Te_2$ }
  \end{minipage}
	\caption{$M_{\rm PV, \it i}^X$ of HOMO, HOMO-1, and HOMO-2
 as a function of dihedral angle $\phi$
as well as the total value ($M_{\rm PV}^X$).
} 
	\label{fig:Mpv_HOMO-2}
\end{figure*}

To understand this spectrum,
a contribution from each orbital is studied.
Figure \ref{fig:Mpv_HOMO-2} shows $M^X_{\rm PV, \it i}$ of HOMO, HOMO-1, and HOMO-2.
It is seen that for almost all dihedral angles
these three contributions are canceled out.
In particular for around $\phi = 90^\circ$,
contributions from HOMO and HOMO-1 are canceled.
In other regions,
the contribution from the HOMO is much smaller than the other contributions
and 
cancellation occurs for contributions from HOMO-1 and HOMO-2.
For any $\phi$, 
the contribution from the HOMO-1 (HOMO-2)
is very large.
The contribution from the HOMO is only large around $\phi = 90^\circ$ 
and in this region, this contribution is much larger than the total contribution.
Any specific relation between the spectrum of $E_{\rm PV}$ in 2a and 2b states
and the contribution from a specific orbital,
while correspondence between the contribution from the HOMO (and HOMO-1) 
and the $E_{\rm PV}$ spectrum in 1a and 1b states
is seen clearly.

In our previous work \cite{Kuroda:2022}, 
it was proposed that the enhancement of $E_{\rm PV}$ by electronic excitation
can be roughly estimated by the $E_{\rm PV}$ contribution from the orbital from which the electron is excited.
In the present work, 
the effectiveness of this estimate is clarified.
The $E_{\rm PV}$ spectrum in 1a and 1b states can be understood in this estimate.
In the 1a and 1b excited states,
the electron in the HOMO is excited to unoccupied orbitals.
Hence, 
if contributions from unoccupied orbitals in the ground state are small in the 1a and 1b states,
the value of $E_{\rm PV}$ in the 1a and 1b states
can roughly be estimated with 
the assumption that the contribution from the HOMO ($M_{\rm PV, HOMO}^X $) is simply lost.
Actually, this rough estimate explains the value of the 1a and 1b states at $\phi = 90^\circ$ \cite{Kuroda:2022}.
(The HOMO contribution in Table XI in Ref.~\cite{Kuroda:2022}
is the contribution from two electrons in the HOMO,
and half of the HOMO contribution is compared to $E_{\rm PV}$ in excited states.)
$E_{\rm PV} $ decreases as angle $\phi$ goes away from $ 90^\circ$,
which corresponds to the decrease of $M_{\rm PV,HOMO}^X $.
On the other hand, this estimate does not predict the value of the 2a and 2b states.
The enhancement in these excited states is not so simple,
since the variation of orbitals from HF orbitals by excitation is important as discussed below.

The enhancement in the first excited state is 
considered to be a special case of enhancement,
and 
cancellation breaking enhancement (CBE) hypothesis is proposed in this paper.
If the following four conditions are satisfied,
significant PVED enhancement is derived in the first excited state
and the enhanced PVED value may be roughly estimated as
the $E_{\rm PV} $ contribution from the HOMO,
\begin{align}
E_{\rm PV} ({\rm CBE})
=
- \sum_n \frac{G_F}{2 \sqrt {2} }  g_V^n M_{\rm PV,HOMO}^n .
\label{eq:CBE}
\end{align}
Since this can be calculated in only HF or DFT computations,
this estimate is considered to be useful.
In more accurate forms, 
it is expressed as
$ - \sum_{n} \sum_{i}^{\rm occ} \frac{G_F}{2 \sqrt {2} } g_V^n r_i^O M_{\rm PV, \it i}^n$
or 
$\frac{G_F}{2 \sqrt {2} } \sum_n g_V^n
( \sum_{i}^{\rm unocc} r_i^U M_{\rm PV, \it i}^n - \sum_{i}^{\rm occ} r_i^O M_{\rm PV, \it i}^n )$,
where $r_i^O$ is 
the excitation ratio of the electron in the $i$-th occupied orbital
and 
$r_i^U$ is the excitation ratio to the $i$-th unoccupied orbital.
These improved estimates require a post-SCF calculation to determine the ratios
and are a much larger cost than the former simple estimate.
The four conditions to be satisfied for deriving significant PVED enhancement are listed below.
\begin{enumerate}
\item The $E_{\rm PV} $ contribution from the HOMO 
is much larger than the total value of $E_{\rm PV} $.
\item In the first excited state,
the electron in the ground state HOMO is dominantly excited.
\item Occupied orbitals in the ground state 
are not significantly modified by the excitation.
\item Contributions from unoccupied orbitals to $E_{\rm PV} $ are small.
\end{enumerate}
The first condition requires that 
the cancellation between contributions is related to the HOMO.
Hence, this condition must be satisfied for enhancement by cancellation breaking.
It is not clear whether this condition is satisfied for most chiral molecules.
However, many molecules, such as 
CHFClBr, alanine, serine, and valine,
satisfy this condition
and this condition can easily be checked by HF or DFT computations.
The second and third conditions are satisfied for many molecules,
at least for vertical excitation of the molecules studied in this work.
If the third condition is not satisfied,
post-SCF computations are necessary 
to determine contributions to $E_{\rm PV} $ from occupied orbitals in excited states.
If the modification of occupied orbitals is not small, 
we can hardly predict the enhancement from the electronic structure in the ground state.
In the first excited state,
excitation is dominated by the electron in the HOMO 
and 
the modification of occupied orbitals is considered to be small.
Hence, the second and third conditions are usually satisfied.
Similar to the first condition, it is not a priori known whether the fourth condition is satisfied.
This condition should be checked for unoccupied orbitals in the ground state.
If unoccupied orbital (to which electrons are excited) contributions are large,
some post-SCF computation, such as CCSD, is needed to determine the contribution.
This hypothesis is related to only the first excited state
and does not comment on the existence of enhancement in any other excited states.
The enhancement mechanism in higher excited states 
is not understood in terms of HF orbitals,
since natural orbitals in excited states are affected by the excitation
and are modified from HF orbitals.
The estimate of the enhancement of Eq.~(\ref{eq:CBE}) 
can be calculated only in SCF computations.
This is highly useful,
and
the validity of the estimate is checked in the following.

\begin{figure*}[p]
\vspace{-10cm}
  \hspace{-1.5cm}
  \begin{minipage}{.46\linewidth}
	\centering
	\includegraphics[width=12cm, bb = 0 0 612 792]{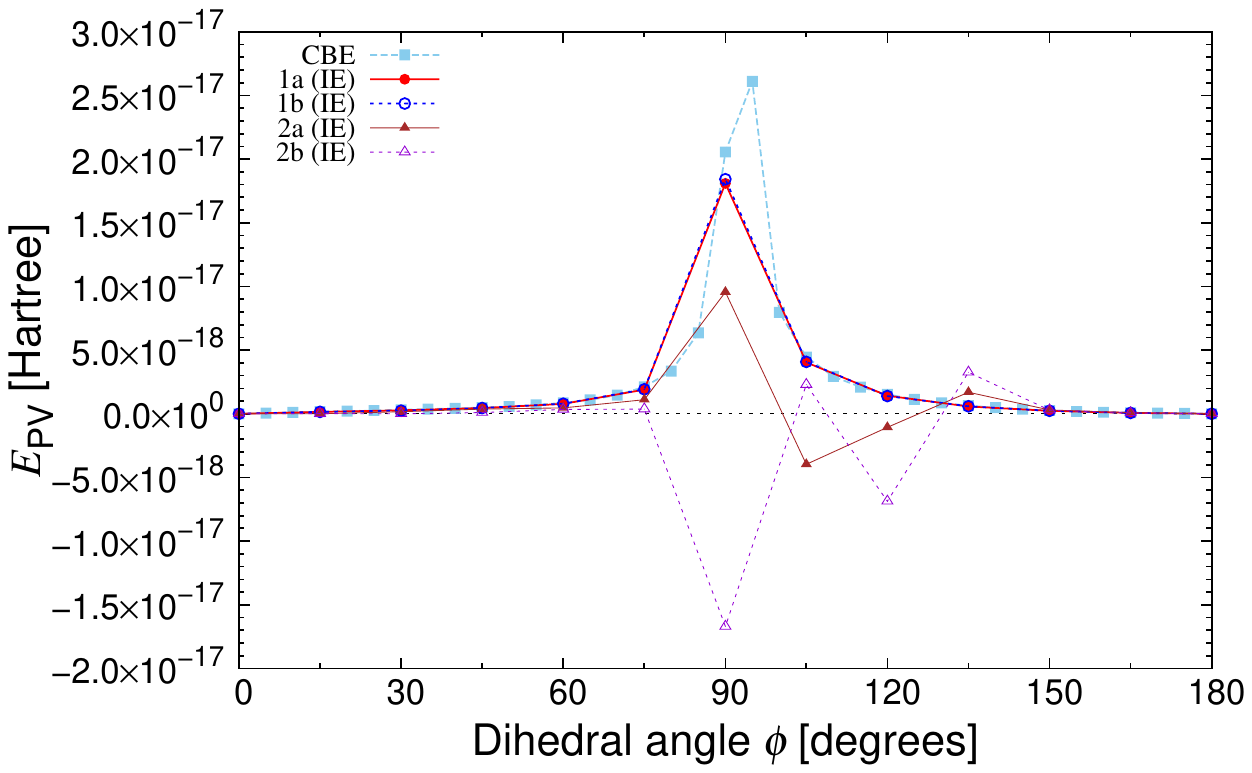} 
 \put(-220,15){ (a) $\rm H_2 O_2$ }
  \end{minipage}
  \hspace{-.5cm}
	\centering
  \begin{minipage}{.46\linewidth}
	\includegraphics[width=12cm, bb = 0 0 612 792]{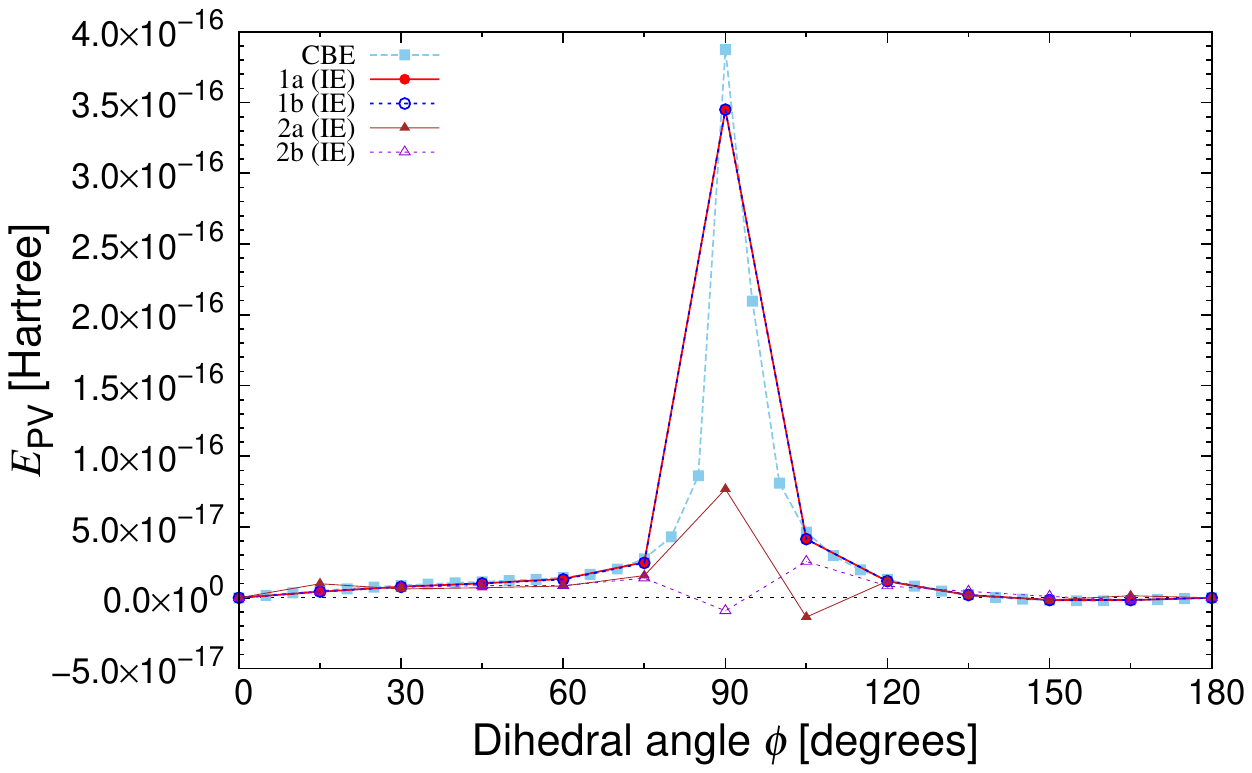} 
 \put(-220,15){ (b) $\rm H_2 S_2$ }
  \end{minipage}
 \\
\vspace{-10cm}
	\centering
  \hspace{-1.5cm}
  \begin{minipage}{.46\linewidth}
	\includegraphics[width=12cm,  bb = 0 0 612 792]{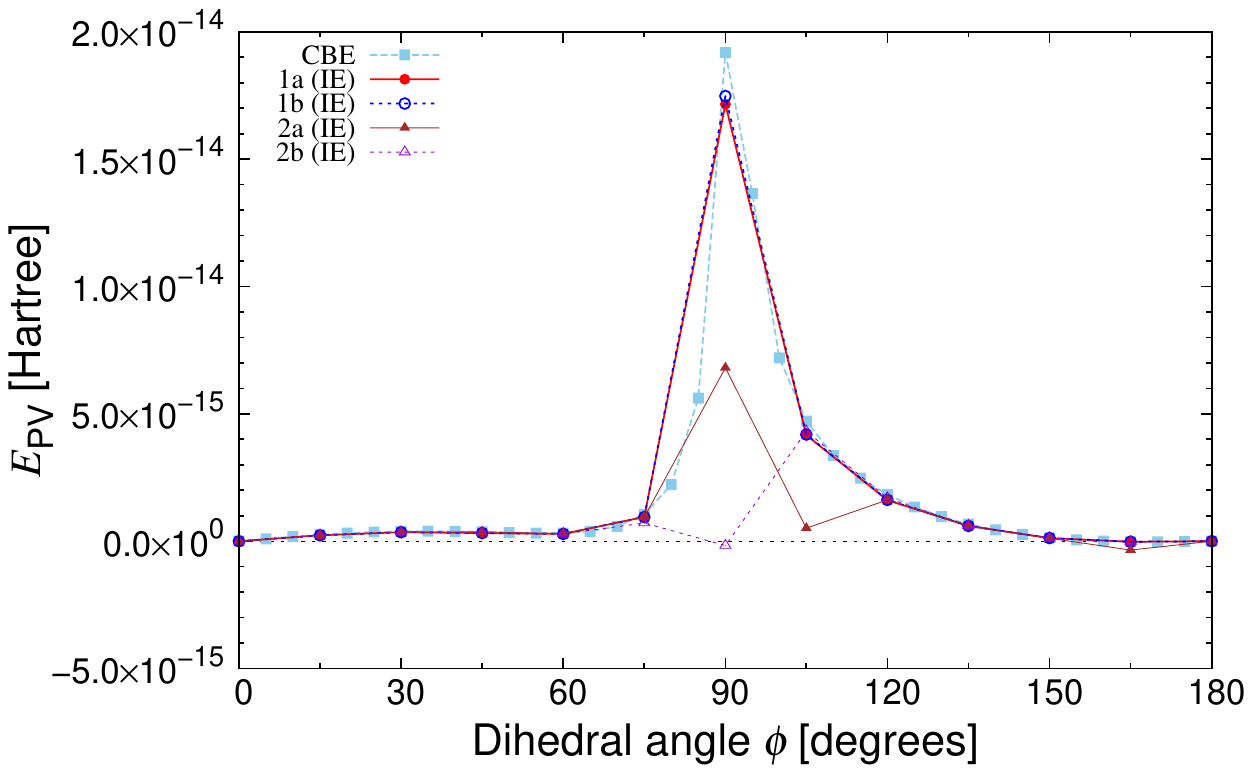} 
 \put(-220,15){ (c) $\rm H_2 Se_2$ }
  \end{minipage}
  \hspace{-.5cm}
	\centering
  \begin{minipage}{.46\linewidth}
	\includegraphics[width=12cm, bb = 0 0 612 792]{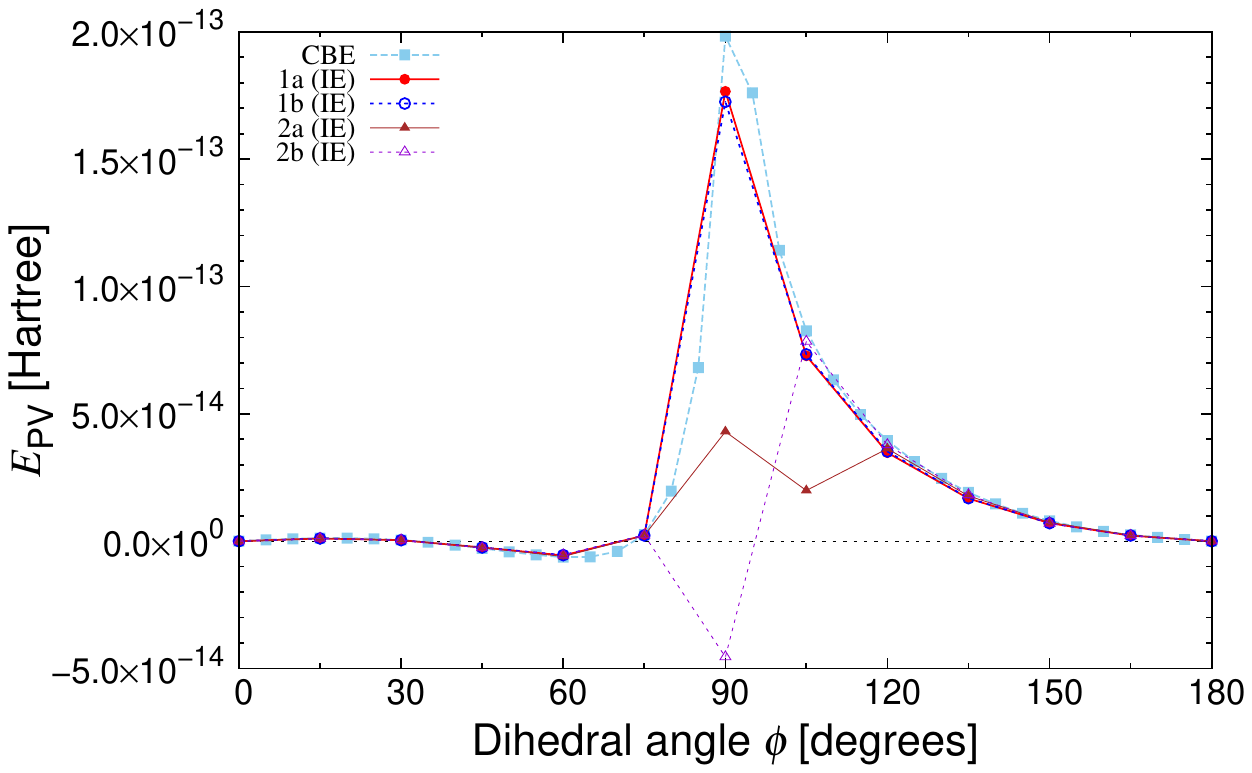} 
 \put(-220,15){ (d) $\rm H_2 Te_2$ }
  \end{minipage}
\vspace{-.5cm}
\caption{Estimate of $E_{\rm PV}$ in the CBE hypothesis
as well as the improved estimate (IE),
$- \sum_{i}^{\rm occ} \frac{G_F}{2 \sqrt {2} } g_V^X r_i^O ( 2 M_{\rm PV, \it i}^X )$.
} 
	\label{fig:Est_Mpv_1}
\vspace{-10cm}
  \hspace{-1.5cm}
  \begin{minipage}{.46\linewidth}
	\centering
	\includegraphics[width=12cm, bb = 0 0 612 792]{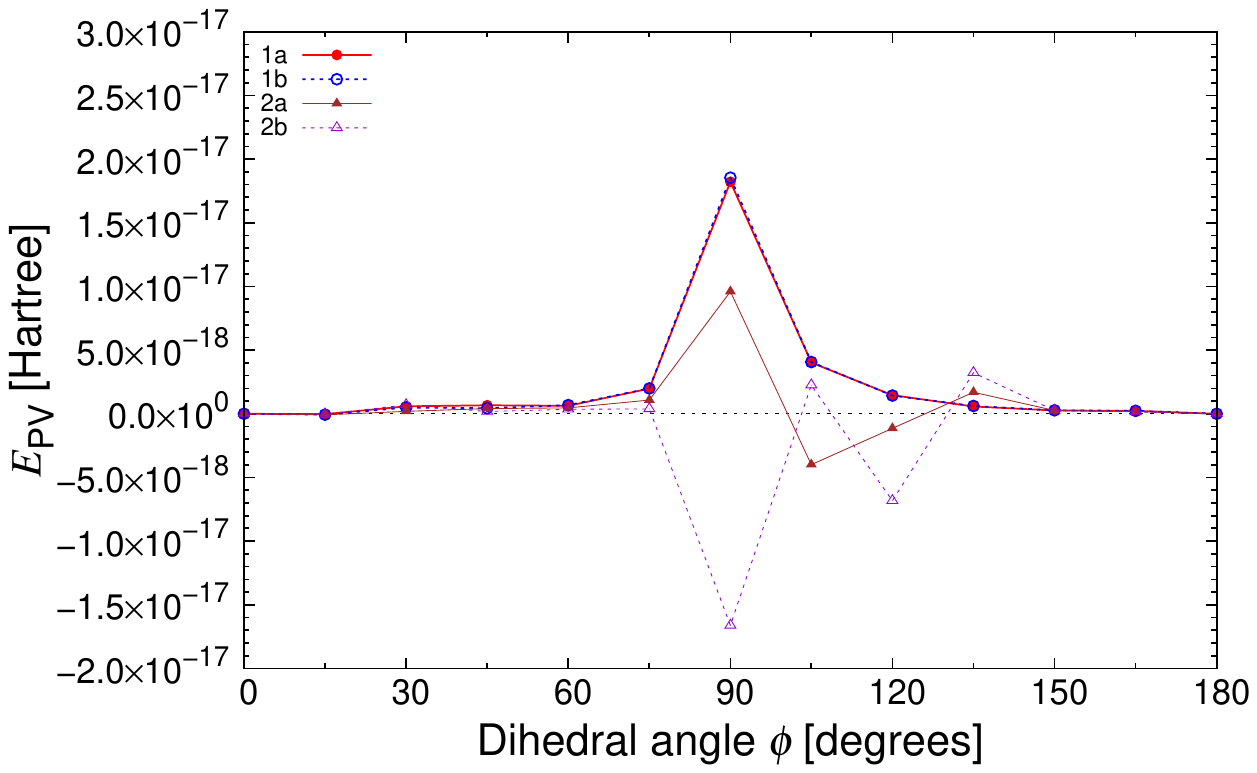} 
 \put(-220,15){ (a) $\rm H_2 O_2$ }
  \end{minipage}
  \hspace{-.5cm}
	\centering
  \begin{minipage}{.46\linewidth}
	\includegraphics[width=12cm, bb = 0 0 612 792]{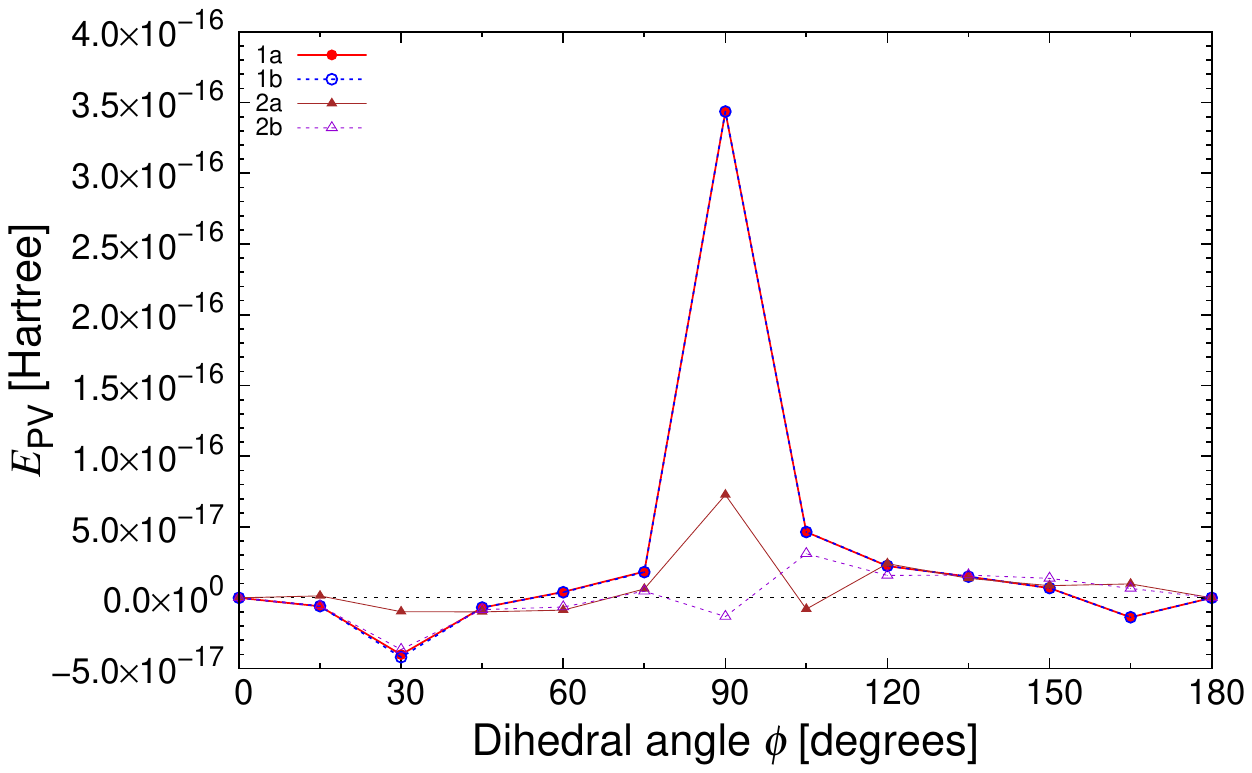} 
 \put(-220,15){ (b) $\rm H_2 S_2$ }
  \end{minipage}
 \\
\vspace{-10cm}
	\centering
  \hspace{-1.5cm}
  \begin{minipage}{.46\linewidth}
	\includegraphics[width=12cm,  bb = 0 0 612 792]{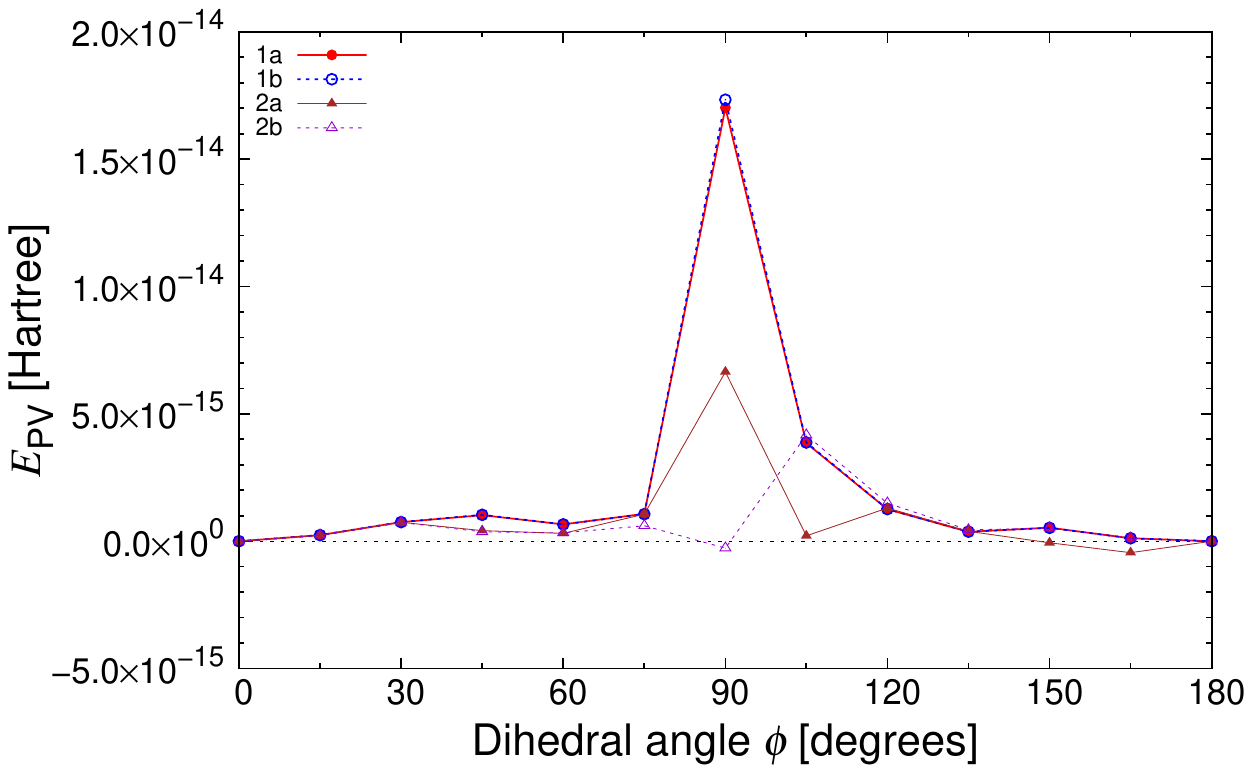} 
 \put(-220,15){ (c) $\rm H_2 Se_2$ }
  \end{minipage}
  \hspace{-.5cm}
	\centering
  \begin{minipage}{.46\linewidth}
	\includegraphics[width=12cm, bb = 0 0 612 792]{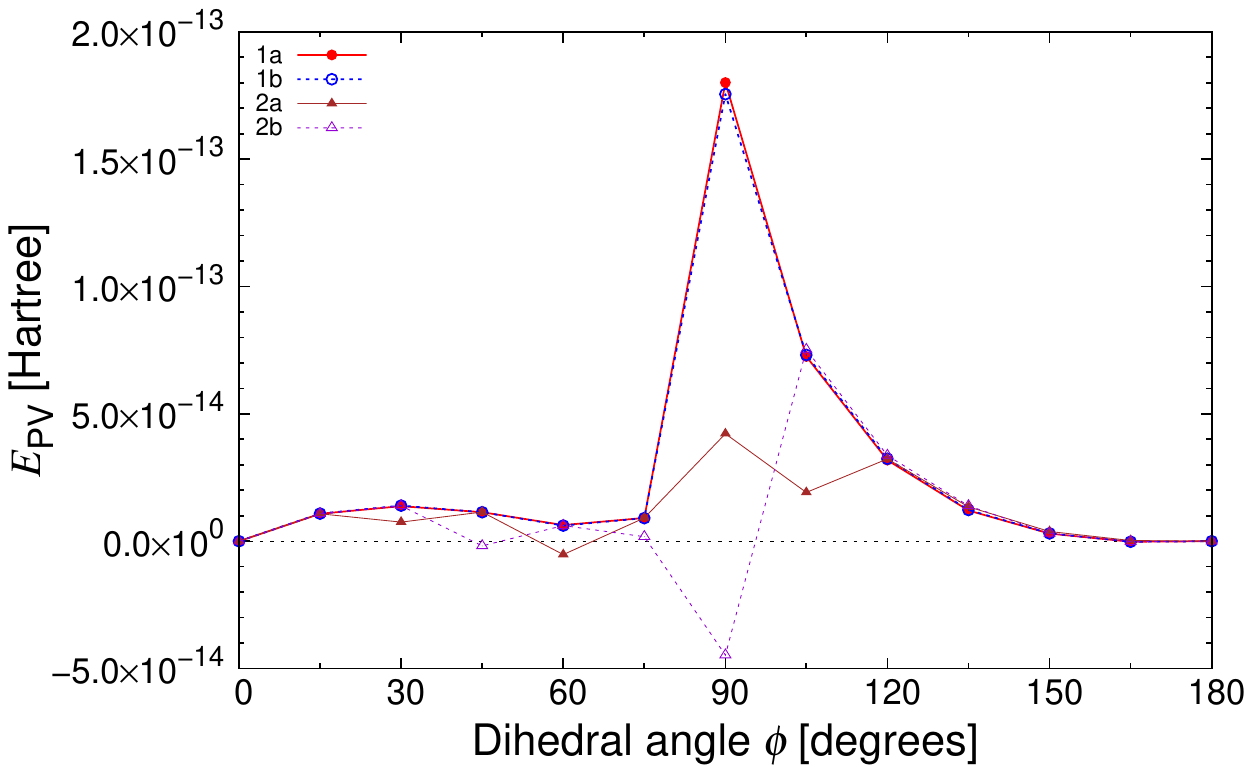} 
 \put(-220,15){ (d) $\rm H_2 Te_2$ }
  \end{minipage}
\vspace{-.5cm}
	\caption{Improved estimate of $E_{\rm PV}$ with contributions from unoccupied orbitals in the ground state,
$\frac{G_F}{2 \sqrt {2} } g_V^X 
( \sum_{i}^{\rm unocc} r_i^U (2 M_{\rm PV, \it i}^X ) - \sum_{i}^{\rm occ} r_i^O (2 M_{\rm PV, \it i}^X ))$.
} 
	\label{fig:Est_Mpv_2}
\end{figure*}

To check the formula evaluating $E_{\rm PV}$ on the CBE hypothesis,
$E_{\rm PV} ({\rm CBE}) $ defined in Eq.~(\ref{eq:CBE})
is shown in Fig.~\ref{fig:Est_Mpv_1}
along with the improved estimate,
$- \sum_{i}^{\rm occ} \frac{G_F}{2 \sqrt {2} } g_V^X r_i^O ( 2 M_{\rm PV, \it i}^X )$,
where the factor 2 in $ 2 M_{\rm PV, \it i}^X $
comes from the existence of two X atoms.
For 1a and 1b states of H$_2X_2$,
$E_{\rm PV} ({\rm CBE}) $ is close to the improved estimate 
since $r_{\rm HOMO}^O \simeq 0.9$.
The distribution curves of the CBE prediction
have the same pattern as $E_{\rm PV}$ of the 1a and 1b states
and the values $E_{\rm PV} ({\rm CBE}) $ at $\phi = 90^\circ$ 
agree accurately with $E_{\rm PV}$ in Fig.~\ref{fig:H2X2_EPV}.
Therefore, the estimate of the CBE hypothesis is confirmed for these molecules.
Small deviations from $E_{\rm PV}$ are seen for heavier elements (Se and Te).
The cause of such deviations is discussed later.
For other excited states,
even the improved estimate cannot correctly predict $E_{\rm PV}$
and the CBE hypothesis is not related to the enhancement in these excited states.
In the 2a and 2b states of H$_2$O$_2$ and H$_2$S$_2$,
$r_{\rm HOMO}^O \lesssim 0.7$ for $\phi = 60-120^\circ$,
and the second condition of the CBE hypothesis is not satisfied.
(To check the CBE hypothesis,
the second condition was studied also for the second excited states.)
In the 2a and 2b states of H$_2$Se$_2$ and H$_2$Te$_2$,
$r_{\rm HOMO}^O \sim 0.9$ except for $\phi \sim 90^\circ$,
where $r_{\rm HOMO}^O \sim 0.2 - 0.4$,
and the second condition of the CBE hypothesis is satisfied
except for $\phi \sim 90^\circ$.

To confirm that the effect of unoccupied orbitals is small,
$\frac{G_F}{2 \sqrt {2} } g_V^X 
( \sum_{i}^{\rm unocc} r_i^U (2 M_{\rm PV, \it i}^X) - \sum_{i}^{\rm occ} r_i^O ( 2 M_{\rm PV, \it i}^X ))$
is shown in Fig.~\ref{fig:Est_Mpv_2}.
In this estimate,  
contributions from virtual orbitals (to which electrons are excited)
are added to the improved estimate in Fig.~\ref{fig:Est_Mpv_1}.
The difference between Figs.~\ref{fig:Est_Mpv_1} and \ref{fig:Est_Mpv_2}
is negligibly small for the 1a and 1b excited state of all H$_2X_2$.
Hence, in these molecules, 
the effect of unoccupied orbitals on the PVED is confirmed to be negligible.
This is just the condition required in the CBE hypothesis.
There is hope that the effect of unoccupied orbitals is small in other molecules as well.
Even in second excited states,
a difference between Figs.~\ref{fig:Est_Mpv_1} and \ref{fig:Est_Mpv_2} is not seen.
Hence, the reason why the CBE hypothesis cannot be applied to the second excited states
is speculated to be because 
the modification of orbitals from HF orbitals in the ground state is not negligible.
That is, the third condition is not satisfied.
In H$_2X_2$, the fourth condition is negligible,
while for other molecules, it is not known whether this condition is satisfied.
In the next section,
CHFClBr-type molecules are taken as test molecules
and the CBE hypothesis is checked.
Before this check, we digress from the CBE hypothesis,
and the electron chirality distribution of each orbital is studied for H$_2X_2$.


\begin{figure*}[tbp]
 \begin{minipage}{.37\linewidth}
 \centering \includegraphics[width=1.00\linewidth]{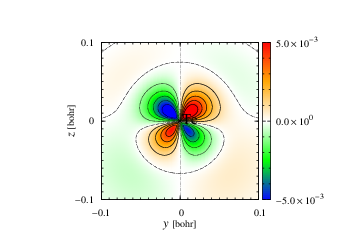}\\
{$\phi= {0}^\circ$}
 \end{minipage}
 \hspace{-40pt}
  \begin{minipage}{.37\linewidth}
 \centering \includegraphics[width=1.00\linewidth]{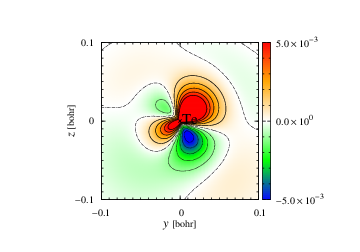}\\
{$\phi= {15}^\circ$}
 \end{minipage} 
 \hspace{-40pt} 
  \begin{minipage}{.37\linewidth}
 \centering \includegraphics[width=1.00\linewidth]{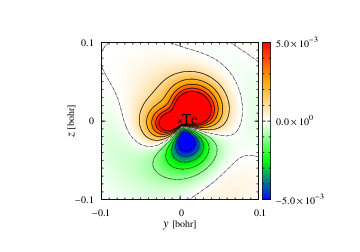}\\
{$\phi= {45}^\circ$}
 \end{minipage} 
 \\
\begin{minipage}{.37\linewidth}
 \centering \includegraphics[width=1.00\linewidth]{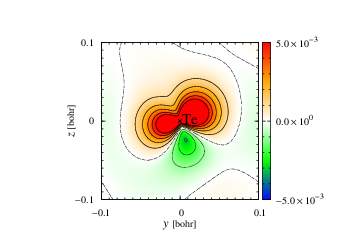}\\
{$\phi= {75}^\circ$}
 \end{minipage}  
 \hspace{-40pt}
  \begin{minipage}{.37\linewidth}
 \centering \includegraphics[width=1.00\linewidth]{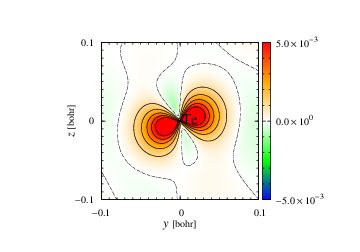}\\
{$\phi= {90}^\circ$}
 \end{minipage}  
 \hspace{-40pt}
  \begin{minipage}{.37\linewidth}
 \centering \includegraphics[width=1.00\linewidth]{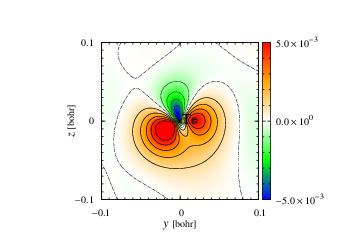}\\
{$\phi= {105}^\circ$}
 \end{minipage}  
 \caption{
Distribution of electron chirality density near a Te nucleus in the ground state of H$_2$Te$_2$.
}
 \label{fig:chiral_dens_H2Te2}
\end{figure*}


\begin{figure*}[tbp]
 \begin{minipage}{.37\linewidth}
 \centering \includegraphics[width=1.00\linewidth]{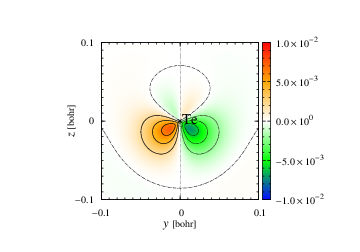}\\
{$\phi= {0}^\circ$}
 \end{minipage}
 \hspace{-40pt} 
  \begin{minipage}{.37\linewidth}
 \centering \includegraphics[width=1.00\linewidth]{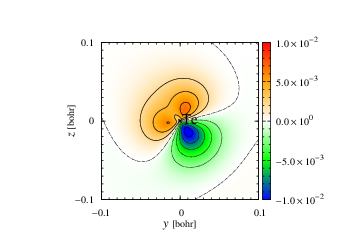}\\
{$\phi= {15}^\circ$}
 \end{minipage} 
 \hspace{-40pt}
  \begin{minipage}{.37\linewidth}
 \centering \includegraphics[width=1.00\linewidth]{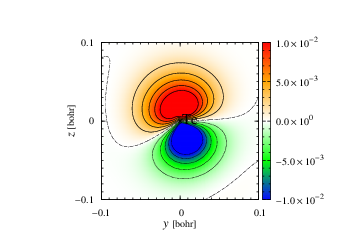}\\
{$\phi= {45}^\circ$}
 \end{minipage} 
 \\
 \begin{minipage}{.37\linewidth}
 \centering \includegraphics[width=1.00\linewidth]{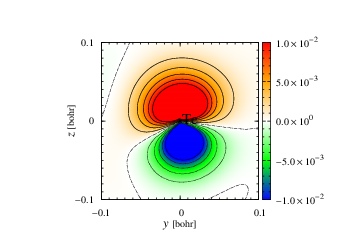}\\
{$\phi= {75}^\circ$}
 \end{minipage}  
 \hspace{-40pt}
  \begin{minipage}{.37\linewidth}
 \centering \includegraphics[width=1.00\linewidth]{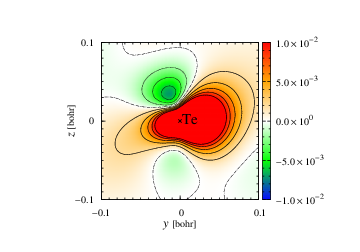}\\
{$\phi= {90}^\circ$}
 \end{minipage}  
 \hspace{-40pt}
  \begin{minipage}{.37\linewidth}
 \centering \includegraphics[width=1.00\linewidth]{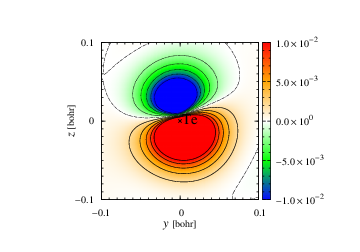}\\
{$\phi= {105}^\circ$}
 \end{minipage}  
 \caption{
Distribution of the HOMO contribution to the electron chirality density near a Te nucleus
in the ground state of H$_2$Te$_2$.
 }
 \label{fig:chiral_dens_HOMO_H2Te2}
\end{figure*}


\begin{figure*}[t]
 \begin{minipage}{.37\linewidth}
 \centering \includegraphics[width=1.00\linewidth]{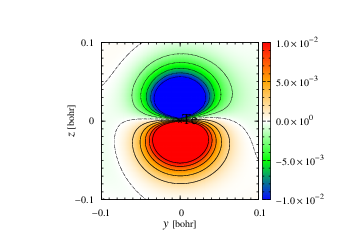}\\
{$\phi= {75}^\circ$}
 \end{minipage}  
 \hspace{-40pt}
  \begin{minipage}{.37\linewidth}
 \centering \includegraphics[width=1.00\linewidth]{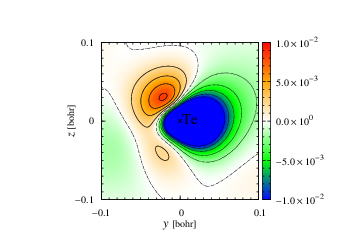}\\
{$\phi= {90}^\circ$}
 \end{minipage}  
 \hspace{-40pt}
  \begin{minipage}{.37\linewidth}
 \centering \includegraphics[width=1.00\linewidth]{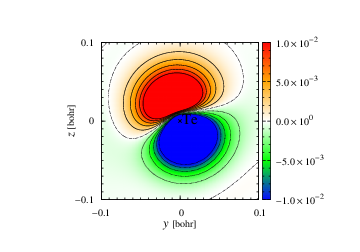}\\
{$\phi= {105}^\circ$}
 \end{minipage}  
 \caption{
Distribution of the HOMO-1 contribution to the electron chirality density near a Te nucleus
in the ground state of H$_2$Te$_2$.
 }
 \label{fig:chiral_dens_HOMO-1_H2Te2}
\end{figure*}

The electron chirality density near a Te nucleus
in the ground state of H$_2$Te$_2$
is shown in Fig.~\ref{fig:chiral_dens_H2Te2}. 
This result is derived by HF computations.
The density is shown in the $yz$ plane,
where the internuclear axis is the $y$ axis.
All atoms are on the $yz$ plane at $\phi = 180^\circ$,
while at $\phi = 0^\circ$,
all atoms are on the $xy$ plane.
Our result is consistent with 
the result in Ref.~\cite{Bast:2011}.
At $\phi= {0}^\circ$, the molecular structure is achiral, 
the electron chirality density is zero at the position of the Te nucleus.
The electron chirality has a $p$-orbital-like distribution 
and positive and negative regions are rotated by $90^\circ$.
The common node point for both regions is located on the Te nucleus.
For larger $\phi$,
the region with positive chirality is extended 
and the electron chirality on the Te nucleus becomes nonzero.
Around $\phi = 90^\circ$, the node of this positive chirality region appears 
on the nucleus,
and $M_{\rm PV}^{\rm Te} \sim 0$.

Figure \ref{fig:chiral_dens_HOMO_H2Te2} shows
the contribution from the HOMO to the electron chirality density near the Te nucleus
in the electronic ground state of H$_2$Te$_2$ by HF computations.
For $\phi= 0-{75}^\circ$, 
the electron chirality density is almost zero at the nucleus position,
which is on the boundary between negative and positive regions.
In comparison between $\phi = 75^\circ$ and $105^\circ$,
the negative and positive regions are reversed.
Around $\phi= {90}^\circ$, 
the transition occurs by the downward transfer of the positive region.
In the contribution from the HOMO-1 (Fig.~\ref{fig:chiral_dens_HOMO-1_H2Te2}),
the region with negative chirality moves downward.
Hence, it is seen that 
a small $E_{\rm PV}$ is realized by 
the cancellation between contributions from HOMO and HOMO-1.
It is speculated that 
in electronic excited states (1a and 1b),
this contribution from the HOMO is removed 
and the cancellation breaking produces large $E_{\rm PV}$.

\subsection{CHFClBr, CHFClI, and CHFBrI}
\label{sec:CHFClbr}

To check the CBE hypothesis for other molecules,
the CHFClBr molecule is studied in this section.
This molecule was actually used in a PVED observational experiment \cite{CHFClBr}.
In addition to CHFClBr,
CHFClI and CHFBrI are also studied.
These two molecules are derived by replacing the Cl or Br atom with an I atom.

\begin{figure}[t]
        \centering
\vspace{-10cm}
  \hspace{-5.5cm}
  \begin{minipage}{.46\linewidth}
\includegraphics[width=12cm, bb = 0 0 612 792]{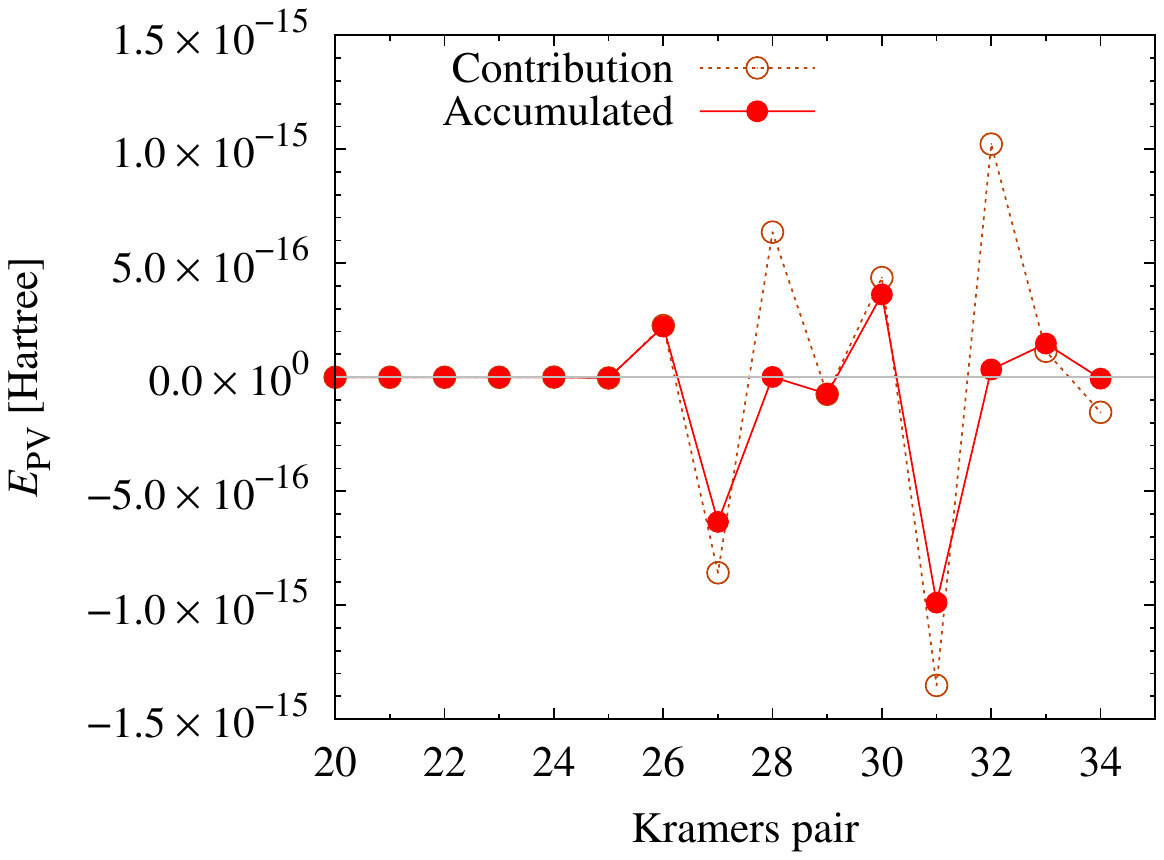} 
 \put(-220,15){ (a) CHFClBr }
  \end{minipage}
 \\
  \hspace{-5.5cm}
  \begin{minipage}{.46\linewidth}
\vspace{-10cm}
\includegraphics[width=12cm, bb = 0 0 612 792]{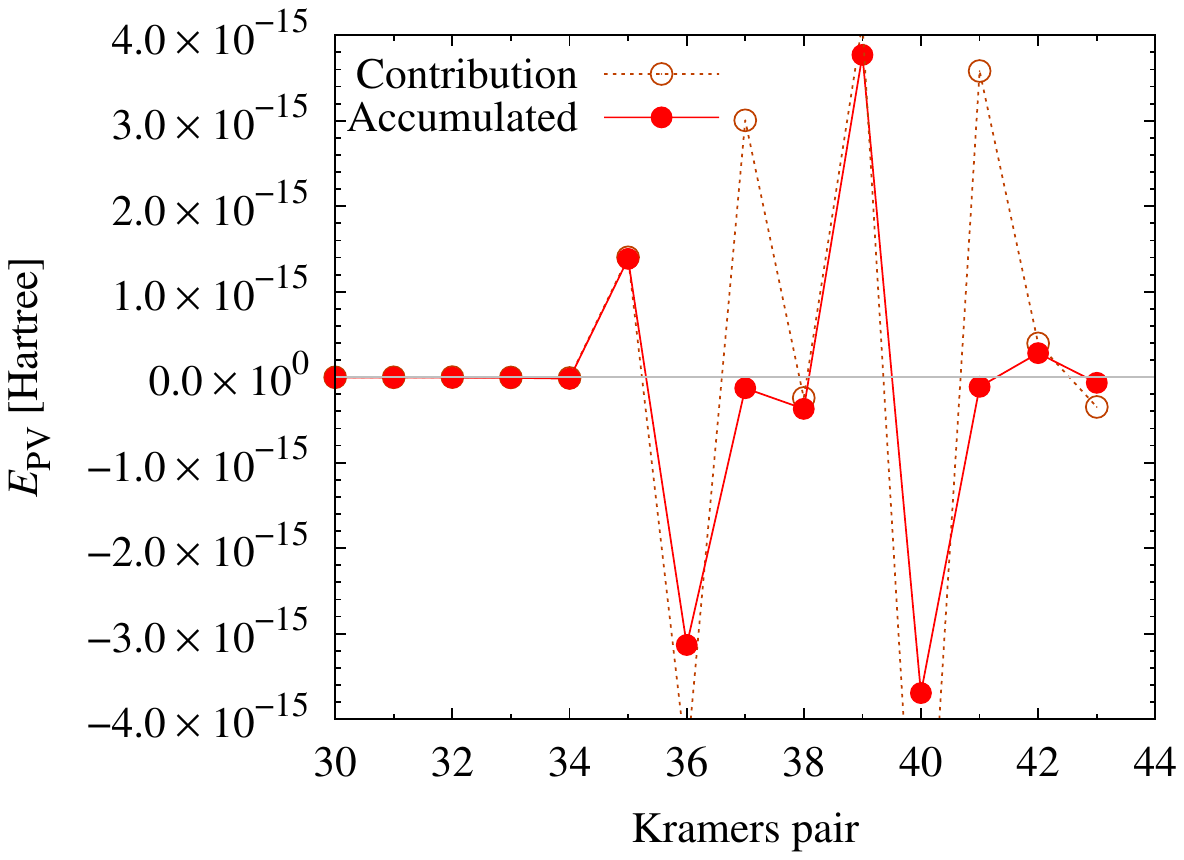} 
 \put(-220,15){\bf (b) CHFClI }
  \end{minipage}
\vspace{-10cm}
 \\
  \hspace{-5.5cm}
  \begin{minipage}{.46\linewidth}
\includegraphics[width=12cm, bb = 0 0 612 792]{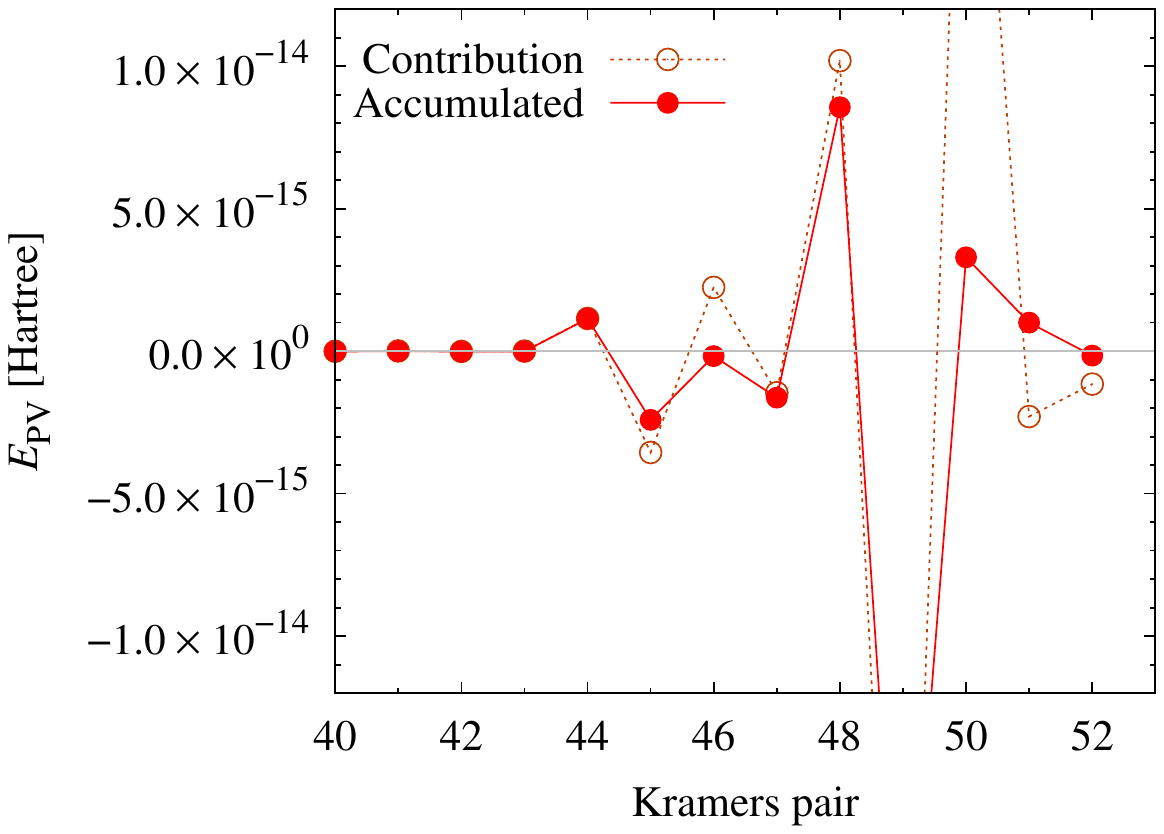} 
 \put(-220,15){\bf (c) CHFBrI}
  \end{minipage}
\vspace{-3mm}
 \\
\caption{
Contribution to $E_{\rm PV} $ from each Kramers pair 
and accumulated value of (a) CHFClBr, (b) CHFClI, and (c) CHFBrI 
in the ground state.
}
         \label{fig:CHFClBr_orbitals}
\end{figure}

Before calculating $E_{\rm PV}$ in excited states,
it should be checked whether these molecules satisfy the conditions of the CBE hypothesis.
First, 
the contribution to $E_{\rm PV} $ from the HOMO 
is confirmed larger by far than the total $E_{\rm PV} $.
In Fig.~\ref{fig:CHFClBr_orbitals},
the contribution to $E_{\rm PV} $ from each Kramers pair in the ground state and accumulated value are shown.
These contributions are computed by the HF method.
In CHFClBr,
the accumulated value of $E_{\rm PV}$ in CHFClBr is $-6.82 \times 10^{-18} $ Hartree,
the contribution from one electron of the HOMO is $-7.71 \times 10^{-17} $ Hartree.
The contribution from the HOMO is about 11 times as large as the total $E_{\rm PV}$.
The first condition of the CBE hypothesis is satisfied for CHFClBr.
For CHFClI and CHFBrI,
the accumulated value of $E_{\rm PV}$ is $-6.64 \times 10^{-17} $ and $-1.56 \times 10^{-16} $ Hartree,
respectively,
while the contribution from the HOMO is $-1.74 \times 10^{-16} $ and $-5.78 \times 10^{-16} $ Hartree,
respectively.
The contribution from the HOMO to $E_{\rm PV}$ 
is also much larger than the accumulated $E_{\rm PV}$ for both molecules,
and the ratio of the HOMO contribution to the total $E_{\rm PV}$ is
about 3 (CHFClI) and 4 (CHFBrI).
These two ratios are smaller than that of CHFClBr,
and 
it is speculated that the enhancement of $E_{\rm PV}$ in CHFClI and CHFBrI is smaller than that in CHFClBr.
Contributions from other valence electrons 
are much larger than the contribution from the HOMO
and even small changes in these orbitals 
may affect the total $E_{\rm PV}$.

The result of CHFClBr in the ground state
is consistent with 
values based on HF computations in previous works \cite{Schwerdtfeger:2005,Thierfelder:2010}.
Contributions to $E_{\rm PV}$ from Kramers pairs
was also reported in Ref.~\cite{Schwerdtfeger:2005}.
The trend in the previous work is the same as observed in our result,
while the values of HOMO and HOMO-1 are significantly smaller than our values.
Their value is about $1/3$ compared to our result.
Nevertheless, 
even their small HOMO value is much larger than the total $E_{\rm PV}$.

The second and third conditions are considered to be satisfied
and, for the first excited state, 
we confirmed that electron excitation dominantly comes from the HOMO in our EOM-CCSD computations.
Hence, the fourth condition that the matrix element of $\sum_n g_V^n M_{\rm PV, \it i}^n$
is small for unoccupied orbitals, is checked.
For the first and second excited states,
contributions to $E_{\rm PV} $ from unoccupied orbitals are much smaller
than that from the HOMO.
Most contributions are less than $1/10$ of the HOMO contribution.
Only the lowest unoccupied molecular orbital (LUMO) contributions in the first and second excited states of CHFClI 
are larger than $1/10$ of the HOMO contribution
and about $1/6$ and $1/7$ of the HOMO contribution, respectively.
Therefore, 
the conditions of the CBE hypothesis are satisfied.



\begin{table*}[tb]
\caption{$E_{{\rm {PV}}}$ in the CHFClBr molecule.
In the rightmost column,
the CBE hypothesis prediction is shown
and this computation is based on the contribution from the HOMO in the HF computation.
}
\centering
\scalebox{1.0}{
\begin{tabular}{crrrrr}
\hline \hline
Atom  & $E_{{\rm {PV}}}/E_{h}$~~ &  \multicolumn{3}{c}{$E_{{\rm {PV}}}/E_{h}$}& $E_{{\rm {PV}}}/E_{h}$~~ \tabularnewline
& (Z-vector) ~~&   \multicolumn{3}{c}{(FFPT)} & (HF)\hspace{5mm} \tabularnewline
\cline{3-5} \cline{4-5} \cline{5-5}
& a\hspace{10mm} & a\hspace{10mm} & 1a\hspace{10mm} & 2a\hspace{10mm} &CBE\hspace{5mm} \tabularnewline
\hline
        Br  & $-6.78 \times 10^{-18}$ & $-6.85 \times 10^{-18}$ & $1.10 \times 10^{-16}$ & $-1.28 \times 10^{-16}$ & $3.94 \times 10^{-17}$\tabularnewline
        Cl  & $  2.46 \times 10^{-18}$ & $  2.48 \times 10^{-18}$ & $1.10 \times 10^{-16}$ & $  1.18 \times 10^{-16}$& $4.00 \times 10^{-17}$\tabularnewline
        F   & $-7.00 \times 10^{-19}$ & $-6.96 \times 10^{-19}$ & $-5.47 \times 10^{-18}$ & $-5.47 \times 10^{-18}$& $-2.62 \times 10^{-18}$ \tabularnewline
        C  & $-3.80 \times 10^{-20}$ & $-3.80 \times 10^{-20}$ & $ 1.14 \times 10^{-19}$ & $-8.14 \times 10^{-19}$ & $4.18 \times 10^{-19}$ \tabularnewline
       Total& $-5.05 \times 10^{-18}$ & $-5.10 \times 10^{-18}$ & $2.15 \times 10^{-16}$ & $-1.58 \times 10^{-17}$ & $7.71 \times 10^{-17}$ \tabularnewline
\hline \hline
\end{tabular}
}
\label{tab:CHFClBr_atom_excited}
\caption{$E_{{\rm {PV}}}$ in the CHFClI molecule.
In the rightmost column,
the CBE hypothesis prediction is shown
and this computation is based on the contribution from the HOMO in the HF computation.
}
\centering
\scalebox{1.0}{
\begin{tabular}{crrrrr}
\hline \hline
Atom  & $E_{{\rm {PV}}}/E_{h}$~~ &  \multicolumn{3}{c}{$E_{{\rm {PV}}}/E_{h}$} & $E_{{\rm {PV}}}/E_{h}$~~\tabularnewline
& (Z-vector) ~~&   \multicolumn{3}{c}{(FFPT)}& (HF)\hspace{5mm} \tabularnewline
\cline{3-5} \cline{4-5} \cline{5-5}
& a\hspace{10mm} & a\hspace{10mm} & 1a\hspace{10mm} & 2a\hspace{10mm} &CBE\hspace{5mm} \tabularnewline
\hline
        I  & $-4.96 \times 10^{-17}$ & $-4.80 \times 10^{-17}$ & $8.44 \times 10^{-16}$ & $-8.71 \times 10^{-16}$& $1.37 \times 10^{-16}$\tabularnewline
        Cl  & $7.36 \times 10^{-18}$ & $7.32 \times 10^{-18}$ & $7.70 \times 10^{-17}$ & $7.86 \times 10^{-17}$& $3.84 \times 10^{-17}$\tabularnewline
        F   & $-2.15 \times 10^{-18}$ & $-2.12 \times 10^{-18}$ & $-4.54 \times 10^{-18}$ & $-4.46 \times 10^{-18}$& $-2.29 \times 10^{-18}$\tabularnewline
        C  & $-1.25\times 10^{-19}$ & $-1.25 \times 10^{-19}$ & $ 4.53 \times 10^{-19}$ & $-2.40 \times 10^{-19}$& $3.50 \times 10^{-19}$\tabularnewline
        Total& $-4.45 \times 10^{-17}$ & $-4.32 \times 10^{-17}$ & $9.16 \times 10^{-16}$ & $-8.02 \times 10^{-16}$& $1.74 \times 10^{-16}$\tabularnewline
\hline \hline
    \end{tabular}
}
\label{tab:CHFClI_atom_excited}
\caption{
$E_{{\rm {PV}}}$ in the CHFBrI molecule.
In the rightmost column, 
the CBE hypothesis prediction is shown
and this computation is based on the contribution from the HOMO in the HF computation.
}
\centering
\scalebox{1.0}{
\begin{tabular}{crrrrr}
\hline \hline
Atom  & $E_{{\rm {PV}}}/E_{h}$~~ &  \multicolumn{3}{c}{$E_{{\rm {PV}}}/E_{h}$} & $E_{{\rm {PV}}}/E_{h}$~~ \tabularnewline
& (Z-vector) ~~&   \multicolumn{3}{c}{(FFPT)} & (HF)\hspace{5mm} \tabularnewline
\cline{3-5} \cline{4-5} \cline{5-5}
& a\hspace{10mm} & a\hspace{10mm} & 1a\hspace{10mm} & 2a\hspace{10mm} & CBE\hspace{5mm} \tabularnewline
\hline
        I  & $-1.91 \times 10^{-16}$ & $-1.84 \times 10^{-16}$ & $-2.04 \times 10^{-16}$ & $1.19 \times 10^{-16}$ & $2.12 \times 10^{-16}$  \tabularnewline
        Br  & $9.56 \times 10^{-17}$ & $9.54 \times 10^{-17}$ & $1.06 \times 10^{-15}$ & $1.15 \times 10^{-15}$& $3.67 \times 10^{-16}$\tabularnewline
        F   & $-1.50 \times 10^{-18}$ & \multicolumn{1}{c}{---}  & \multicolumn{1}{c}{---} & \multicolumn{1}{c}{---}& $-1.21 \times 10^{-18}$ \tabularnewline
        C  & $-9.42\times 10^{-20}$ & $-9.45 \times 10^{-20}$ & $ -2.25 \times 10^{-19}$ & $-6.46 \times 10^{-19}$& $2.18 \times 10^{-19}$\tabularnewline
        Total& $-9.67 \times 10^{-17}$ & $-8.91 \times 10^{-17}$ & $8.60 \times 10^{-16}$ & $1.26 \times 10^{-15}$& $5.78 \times 10^{-16}$\tabularnewline
\hline \hline
    \end{tabular}
}
\label{tab:CHFBrI_atom_excited}
\end{table*}

Since the conditions of the CBE hypothesis are satisfied,
the PVED enhancement is studied for these molecules.
The PVED in excited states calculated by the EOM-CCSD method
is shown in Tables \ref{tab:CHFClBr_atom_excited}, \ref{tab:CHFClI_atom_excited},
and \ref{tab:CHFBrI_atom_excited}.
The adopted values of $\lambda^n$ are
$\lambda^{\rm C} = 10^{-1}$ (a.u.),
$\lambda^{\rm F} = 10^{-3}$ (a.u.),
$\lambda^{\rm Cl} = 10^{-3}$ (a.u.),
$\lambda^{\rm Br} = 10^{-3}$ (a.u.), and
$\lambda^{\rm I} = 10^{-5}$ (a.u.).
In the result of CHFBrI, 
the contribution of the F atom was not included,
since the EOM-CCSD computation with the F atom perturbation did not converge.
Since contributions from the F atom are subdominant in Z-vector calculations of both CHFClBr and CHFClI,
the neglected contribution from the F atom can safely be dropped in CHFBrI.
For all molecules, 
the CBE enhancement of $ E_{\rm PV} $ is observed in the first excited state,
and the enhancement ratios are
-42 (CHFClBr), -21 (CHFClI), and -9 (CHFBrI).
The observed enhancement of these molecules is weaker than H$_2X_2$ at $\phi = 90^\circ $
and it is attributed to the weaker dominance of the HOMO contributions to $E_{\rm PV}$.
Molecules with an I atom have a smaller HOMO contribution ratio to the total $E_{\rm PV}$,
and the enhancement ratio is small.
The heaviest element has the largest contribution in the ground state, 
while in excited states this is not always true.
In excited states of CHFClBr,
the contribution from the Cl atom is comparable to that from the Br atom.
In excited states of CHFBrI,
the contribution from the Br atom is larger than that from the I atom.
For CHFClBr,   
the enhancement in the first excited state 
is much larger than the second excited state,
while for CHFClI and CHFBrI 
the enhancement of two excited states is comparable.
The smallness of the second excited state in CHFClBr
is due to the cancellation between the contributions of Br and Cl atoms.
The first and second excited states of these three molecules are in the same spin triplet.
The excitation energies of the first and second excited states are
4.960 and 4.961 eV (CHFClBr),
3.667 and 3.667 eV (CHFClI),
and 3.577 and 3.578 eV (CHFBrI), respectively.

\begin{figure}[t]
        \centering
\vspace{-10cm}
\hspace{-5cm}
  \begin{minipage}{.46\linewidth}
\includegraphics[width=12cm, bb = 0 0 612 792]{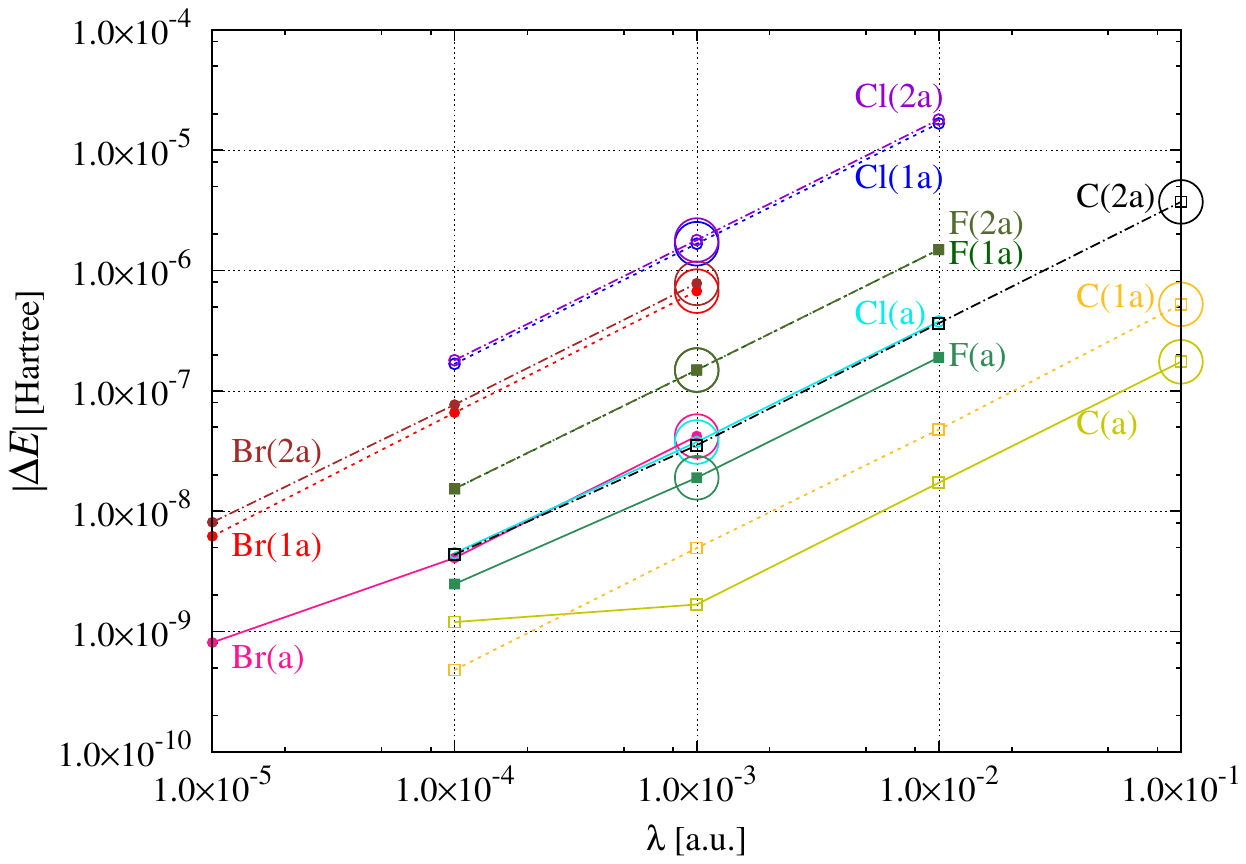} 
 \put(-230,15){ (a) CHFClBr }
  \end{minipage}
 \\
\vspace{3mm}
\vspace{-10cm}
\hspace{-5cm}
  \begin{minipage}{.46\linewidth}
\includegraphics[width=12cm, bb = 0 0 612 792]{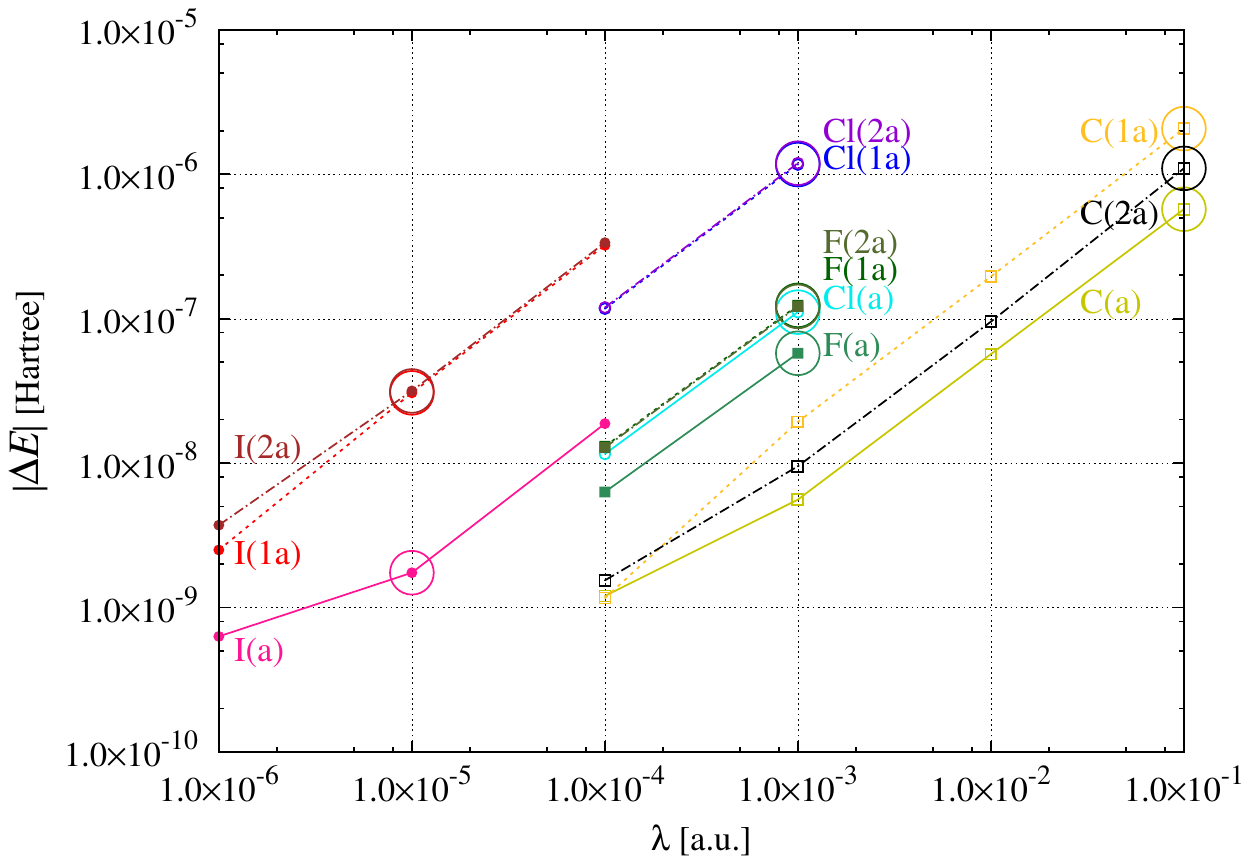} 
 \put(-230,15){\bf (b) CHFClI }
  \end{minipage}
\vspace{3mm}
 \\
\caption{
Dependence of the energy difference between $\pm \lambda^n $ 
on $\lambda^n$.
Adopted values of $\lambda^n$ are encircled. }
\label{fig:deltaE}
\end{figure}


\begin{table*}[tb]
\caption{Dependence of $E_{{\rm {PV}}}$ on perturbation parameter $\lambda$ within the FFPT method in
the ground (a), first excited (1a), and second excited (2a) electronic states of the CHFClBr molecule.
Adopted values are emphasized by italic letters.
}
\centering
\scalebox{1.0}{
\begin{tabular}{ccrrr}
\hline \hline
Atom  & $\lambda$ &  \multicolumn{3}{c}{$E_{{\rm {PV}}}/E_{h}$(FFPT)}\tabularnewline
\cline{3-5} \cline{4-5} \cline{5-5}
& (a.u.) & a\hspace{10mm} & 1a\hspace{10mm} & 2a\hspace{10mm}  \tabularnewline
\hline
        Br  & $\mathit{1.0 \times 10^{-3}}$ & $\mathit{-6.85 \times 10^{-18}}$ & $\mathit{1.10 \times 10^{-16}}$ & $\mathit{-1.28 \times 10^{-16}}$ \tabularnewline
            & $1.0 \times 10^{-4}$ & $-6.66 \times 10^{-18}$ & $1.07 \times 10^{-16}$ & $-1.25 \times 10^{-16}$ \tabularnewline    
            & $1.0 \times 10^{-5}$ & $-1.32 \times 10^{-17}$ & $1.01 \times 10^{-16}$ & $-1.32 \times 10^{-16}$ \tabularnewline 
\hline                           
        Cl  & $1.0 \times 10^{-2}$ & $  2.48 \times 10^{-18}$ & $1.10 \times 10^{-16}$ & $  1.18 \times 10^{-16}$\tabularnewline
            & $\mathit{1.0 \times 10^{-3}}$ & $\mathit{2.48 \times 10^{-18}}$ & $\mathit{1.10 \times 10^{-16}}$ & $\mathit{1.18 \times 10^{-16}}$\tabularnewline
            & $1.0 \times 10^{-4}$ & $  2.96 \times 10^{-18}$ & $1.10 \times 10^{-16}$ & $  1.18 \times 10^{-16}$\tabularnewline 
\hline                                     
        F   & $1.0 \times 10^{-2}$ & $-6.96 \times 10^{-19}$ & $-5.47 \times 10^{-18}$ & $-5.48 \times 10^{-18}$\tabularnewline
             & $\mathit{1.0 \times 10^{-3}}$ & $\mathit{-6.96 \times 10^{-19}}$ & $\mathit{-5.47 \times 10^{-18}}$ & $\mathit{-5.47 \times 10^{-18}}$\tabularnewline
             & $1.0 \times 10^{-4}$ & $-9.11 \times 10^{-19}$ & $-5.64 \times 10^{-18}$ & $-5.65 \times 10^{-18}$\tabularnewline   
\hline                                    
        C  & $\mathit{1.0 \times 10^{-1}}$ & $\mathit{-3.80 \times 10^{-20}}$ & $\mathit{1.14 \times 10^{-19}}$ & $\mathit{-8.14 \times 10^{-19}}$\tabularnewline
            & $1.0 \times 10^{-2}$ & $-3.76 \times 10^{-20}$ & $ 1.04 \times 10^{-19}$ & $-7.98 \times 10^{-19}$\tabularnewline
            & $1.0 \times 10^{-3}$ & $-3.67 \times 10^{-20}$ & $ 1.08 \times 10^{-19}$ & $-7.73 \times 10^{-19}$\tabularnewline    
            & $1.0 \times 10^{-4}$ & $-2.62 \times 10^{-20}$ & $ 1.05 \times 10^{-19}$ & $-9.53 \times 10^{-19}$\tabularnewline                                    
\hline \hline
    \end{tabular}
}
\label{tab:CHFClBr_lambda}
\end{table*}


\begin{table*}[tb]
\caption{Dependence of $E_{{\rm {PV}}}$ on perturbation parameter $\lambda$ within the FFPT method in
the ground (a), first excited (1a), and second excited (2a) electronic states of the CHFClI molecule.
Adopted values are emphasized by italic letters.
}
\centering
\scalebox{1.0}{
\begin{tabular}{ccrrr}
\hline \hline
Atom  & $\lambda$ &  \multicolumn{3}{c}{$E_{{\rm {PV}}}/E_{h}$(FFPT)}\tabularnewline
\cline{3-5} \cline{4-5} \cline{5-5}
& (a.u.) & a\hspace{10mm} & 1a\hspace{10mm} & 2a\hspace{10mm}  \tabularnewline
\hline
        I   & $1.0 \times 10^{-4}$ & $-5.17 \times 10^{-17}$ & $8.88 \times 10^{-16}$ & $-9.27 \times 10^{-16}$ \tabularnewline    
        & $\mathit{1.0 \times 10^{-5}}$ & $\mathit{-4.80 \times 10^{-17}}$ & $\mathit{8.44 \times 10^{-16}}$ & $\mathit{-8.71 \times 10^{-16}}$ \tabularnewline            
            & $1.0 \times 10^{-6}$  & $-1.74 \times 10^{-16}$ & $6.89 \times 10^{-16}$ & $-1.03 \times 10^{-15}$ \tabularnewline      
\hline                                            
        Cl & $\mathit{1.0 \times 10^{-3}}$ & $\mathit{7.32 \times 10^{-18}}$ & $\mathit{7.70 \times 10^{-17}}$ & $\mathit{7.86 \times 10^{-17}}$\tabularnewline
            & $1.0 \times 10^{-4}$ & $  7.62 \times 10^{-18}$ & $7.72 \times 10^{-17}$ & $  7.89 \times 10^{-17}$\tabularnewline   
\hline                                               
        F   & $\mathit{1.0 \times 10^{-3}}$ & $\mathit{-2.12 \times 10^{-18}}$ & $\mathit{-4.54 \times 10^{-18}}$ & $\mathit{-4.46 \times 10^{-18}}$\tabularnewline
             & $1.0 \times 10^{-4}$ & $-2.32 \times 10^{-18}$ & $-4.79 \times 10^{-18}$ & $-4.71 \times 10^{-18}$\tabularnewline  
\hline                                                  
        C  & $\mathit{1.0 \times 10^{-1}}$ & $\mathit{-1.25 \times 10^{-19}}$ & $\mathit{4.53 \times 10^{-19}}$ & $\mathit{-2.40 \times 10^{-19}}$\tabularnewline
            & $1.0 \times 10^{-2}$ & $-1.24 \times 10^{-19}$ & $ 4.28 \times 10^{-19}$ & $-2.09 \times 10^{-19}$\tabularnewline
            & $1.0 \times 10^{-3}$ & $-1.22 \times 10^{-19}$ & $ 4.22 \times 10^{-19}$ & $-2.08 \times 10^{-19}$\tabularnewline
            & $1.0 \times 10^{-4}$ & $-2.62 \times 10^{-19}$ & $ 2.54 \times 10^{-19}$ & $-3.37 \times 10^{-19}$\tabularnewline                
\hline \hline
    \end{tabular}
}
\label{tab:CHFClI_lambda}
\end{table*}

As noted in Sec.~\ref{sec:comp},
a loose convergence threshold is used
for EOM-CCSD computations of excited states.
Since the energy difference by the perturbation
is $ \lambda^n M_{\rm PV}^n$,
computations should be sufficiently accurate 
to include this effect.
If this effect is correctly included,
the calculated energy difference,
$\Delta E^n \equiv E^n( \lambda^n) - E^n (-\lambda^n) $,
is proportional to the value of $\lambda^n$
and $E_{\rm PV}$ is then independent of $\lambda^n$.
To check this, 
$\Delta E^n $ of CHFClBr and CHFClI are shown in Fig.~\ref{fig:deltaE}.
For CHFBrI, $ E^n$ are calculated in a single $\lambda^n$ value speculated 
from computations of CHFClBr and CHFClI
to save computational costs.
For our chosen values of $\lambda^n$,
all $\Delta E^n$ are proportional to $\lambda^n$.
This result clearly confirms that
our loose convergence computations are sufficiently accurate
to derive $E_{\rm PV}$.
Tables \ref{tab:CHFClBr_lambda} and \ref{tab:CHFClI_lambda} show 
that the calculated $E_{\rm PV}$ is independent of $\lambda^n$
around our adopted value of $\lambda^n$.

Since the conditions of the CBE hypothesis are satisfied
and the enhancement of $ E_{\rm PV}$ is confirmed,
the accuracy of the simple estimate of $E_{\rm PV}$ by the CBE hypothesis should be checked.
In Tables \ref{tab:CHFClBr_atom_excited}, \ref{tab:CHFClI_atom_excited},
and \ref{tab:CHFBrI_atom_excited},
the prediction of $E_{\rm PV}$ based on the CBE hypothesis is also shown.
The CBE hypothesis can successfully predict the existence of enhancement 
and estimate the order of magnitude of $ E_{\rm PV}$.
However, its accuracy is significantly worse compared to that calculated for H$_2X_2$.
The predicted values of $E_{\rm PV}$
are $1/3$, $1/5$, and $2/3$ of the values derived by EOM-CCSD
for CHFClBr, CHFClI, and CHFBrI in the first excited state, respectively.
This gap is not narrowed significantly, even with the usage of the improved formulas.
Hence, the absence of effects of other HF orbitals in the CBE prediction
is not an important factor of this inaccuracy,
and the effect of the modification of some occupied orbitals
is speculated to be important.
The contributions from HOMO-2 and HOMO-3 are 
much larger than the HOMO contribution as seen in Fig.~\ref{fig:CHFClBr_orbitals}.
Even the small change of an orbital with a large contribution
may give a large change of $E_{\rm PV}$.
In other words, 
if the contribution of the HOMO is not a dominant part of $E_{\rm PV}$,
the accuracy of the CBE prediction (Eq.~(\ref{eq:CBE}))
may be insufficient
despite the successful prediction of the existence of enhancement.

Lastly, the dependence on computational conditions is studied using CHFClBr as an example.
The value of $E_{\rm PV}$ in the ground state is strongly dependent on the choice of basis sets.
In the CCSD computation with aug-cc-pVTZ, cc-pVTZ, and aug-cc-pVDZ basis sets \cite{ccpv_basis_set},
$E_{\rm PV} = -2.54 \times 10^{-18}, -1.83 \times 10^{-18}$, and $-1.60 \times 10^{-18}$ Hartree.
In these computations, basis sets are used in the uncontracted form.
These results with basis sets based on nonrelativistic theory
are much less than our results.
For CCSD computations with higher virtual cutoff ($200$ Hartree) of CCSD computations,
our results do significantly not change and the difference is negligible
in both ground and excited states.
When we use extended correlating orbitals ($3p3d4s4p$(Br), $2p3s3p$(Cl), $2s2p$(F), $2s2p$(C), and $1s$(H)),
the change of $ E_{\rm PV} $ is up to 10\% 
in the ground and first excited states,
while in the second excited state
$ E_{\rm PV} $ increases as $E_{\rm PV}= -3.54\times 10^{-17}$ Hartree.
The usage of more extension of correlating orbitals may be important
and due to the limit of our computational resources
our correlating orbitals are restricted to valence electrons.
Nevertheless, this is sufficiently accurate for the confirmation of the existence of the enhancement.



\section{Conclusion}

In this paper, 
we have studied the $E_{\rm PV}$ enhancement in electronic excited states
for H$_2X_2$ ($X =$ O, S, Se, Te), CHFClBr, CHFClI, and CHFBrI.
The enhancement of $E_{\rm PV}$ of H$_2X_2$ in excited states
is highly dependent on the dihedral angle.
Through the study of the dihedral angle dependence,
the CBE hypothesis has been proposed.
This hypothesis has been confirmed for excited states of CHFClBr, CHFClI, and CHFBrI.
If the contribution from the HOMO to $E_{\rm PV}$ in the ground state
is larger than the sum of those from all occupied orbitals,
the CBE hypothesis predicts the existence of the PVED enhancement 
in the first excited state.
Moreover, if the contribution from the HOMO to $E_{\rm PV}$ 
dominates over other contributions,
the CBE hypothesis can predict the value of $E_{\rm PV}$ in the first excited state.

We believe that 
PVED enhancement by the CBE hypothesis
occurs for many other chiral molecules.
This enhancement mechanism may be a hint for choosing
a chiral molecule for future experiments used to observe PVED.
For example, 
the measurement of the vibrational frequency difference 
between two enantiomers of a chiral molecule in excited states
may be a candidate for future experiments.
The precise measurement of the excitation energy difference between an enantiomeric pair 
is also confirmation of the existence of PVED.
In this paper, 
molecular structures of CHFClBr, CHFClI, and CHFBrI in excited states
are chosen to be the same as the ground state.
Nevertheless, these molecules are not considered to be achiral in excited states.
In optimized structures,
the PVED may not be enhanced significantly,
but we expect that large PVED enhancement is realized in the structures.
The investigation of the PVED in the optimized structure of an excited state
is important for actual experimental planning.
For a molecule whose structure in the first excited state
is not so different from that in the ground state,
the PVED enhancement in the CBE hypothesis 
is available for future experiments.\\



\begin{acknowledgments}
This work was supported by Grants-in-Aid for Scientific Research (17K04982, 21H00072, and 22K12060).
A.S. acknowledges financial support from the Japan Society for the Promotion of Science (JSPS) KAKENHI Grant No. 20K22553 and 21K14643.
The work of N. K. was supported by JST SPRING, Grant Number JPMJSP2110.
In this research work we used the supercomputer of ACCMS, Kyoto University.
\end{acknowledgments}

\end{document}